\documentclass[pdflatex,sn-mathphys-num]{sn-jnl}

\usepackage{graphicx}%
\usepackage{multirow}%
\usepackage{amsmath,amssymb,amsfonts}%
\usepackage{amsthm}%
\usepackage{mathrsfs}%
\usepackage[title]{appendix}%
\usepackage{xcolor}%
\usepackage{textcomp}%
\usepackage{manyfoot}%
\usepackage{booktabs}%
\usepackage{algorithm}%
\usepackage{algorithmicx}%
\usepackage{algpseudocode}%
\usepackage{listings}%

\theoremstyle{thmstyleone}%
\newtheorem{theorem}{Theorem}
\newtheorem{proposition}[theorem]{Proposition}%

\theoremstyle{thmstyletwo}%
\newtheorem{example}{Example}%
\newtheorem{remark}{Remark}%

\theoremstyle{thmstylethree}%
\newtheorem{definition}{Definition}%

\begin{document}

\title[Article Title]{Radiomics in Medical Imaging: Methods, Applications, and Challenges}


\author*[1]{\fnm{Fnu} \sur{ Neha}}\email{neha@kent.edu}

\author[1]{\fnm{Deepak Kumar} \sur{Shukla}}\email{ds1640@scarletmail.rutgers.edu}
\equalcont{This author contributed equally to this work.}


\affil*[1]{\orgdiv{Department of Computer Science}, \orgname{Kent State University}, \orgaddress{\street{Kent}, \city{Ohio}, \postcode{44240}, \country{USA}}}

\affil[1]{\orgdiv{Rutgers Business School}, \orgname{Rutgers University}, \orgaddress{\city{Newark}, \state{New Jersey}, \postcode{07102}, \country{USA}}}

\abstract{Radiomics enables quantitative medical image analysis by converting imaging data into structured, high-dimensional feature representations for predictive modeling. Despite methodological developments and encouraging retrospective results, radiomics continue to face persistent challenges related to feature instability, limited reproducibility, validation bias, and restricted clinical translation. Existing reviews largely focus on application-specific outcomes or isolated pipeline components, with limited analysis of how interdependent design choices across acquisition, preprocessing, feature engineering, modeling, and evaluation collectively affect robustness and generalizability. This survey provides an end-to-end analysis of radiomics pipelines, examining how methodological decisions at each stage influence feature stability, model reliability, and translational validity. This paper reviews radiomic feature extraction, selection, and dimensionality reduction strategies; classical machine and deep learning–based modeling approaches; and ensemble and hybrid frameworks, with emphasis on validation protocols, data leakage prevention, and statistical reliability. Clinical applications are discussed with a focus on evaluation rigor rather than reported performance metrics. The survey identifies open challenges in standardization, domain shift, and clinical deployment, and outlines future directions such as hybrid radiomics–artificial intelligence models, multimodal fusion, federated learning, and standardized benchmarking.
}




\maketitle

\section{Introduction}\label{sec1}
Medical imaging plays an important role in clinical decision-making, supporting diagnosis, prognosis, treatment planning, and disease monitoring. Advances in medical imaging have improved spatial resolution and increased availability across routine clinical workflows; however, clinical interpretation remains predominantly qualitative, subject to inter-observer variability and limited sensitivity to subtle phenotypic patterns embedded in high-dimensional image data \cite{ma2025towards}. These limitations highlight the need for quantitative image-derived biomarkers that complement visual interpretation and provide reproducible clinical evidence.

Radiomics addresses this need by enabling high-throughput extraction of quantitative descriptors from medical images, transforming visual information into structured, analyzable data \cite{gillies2016radiomics}. Radiomic features characterize tissue morphology, intensity distributions, and spatial heterogeneity in a non-invasive manner and have been explored across diverse imaging modalities, including CT, MRI, PET, ultrasound, and digital pathology \cite{sala2017unravelling}. Through systematic quantification of imaging phenotypes, radiomics supports data-driven disease characterization and precision medicine approaches.

Radiomics has been shown clinical relevance across multiple application domains such as in oncology, radiomic signatures have been analyzed for tumor characterization, subtype differentiation, grading, outcome prediction, and therapy response assessment \cite{liu2019applications}. Applications have also extended to neurology and cardiology, where quantitative imaging phenotypes contribute to assessment of disease progression and functional outcomes beyond visually apparent patterns\cite{zhou2018radiomics, oikonomou2020artificial}.

Despite growing interest, several methodological challenges, including variability in image acquisition, preprocessing, segmentation, and feature computation, limit the robustness and clinical translation of current radiomics research. High-dimensional feature spaces combined with limited cohort sizes increase the risk of overfitting and compromise reproducibility. Furthermore, external validation and multi-center evaluations remain insufficient, reducing confidence in model generalizability \cite{castillo2021multi}.

Existing surveys examine specific radiomics applications or workflow components, such as feature engineering or outcome prediction, and present an end-to-end view of the radiomics pipeline \cite{vial2018role, linton2025radiomics, xu2025addressing, perniciano2025insights, zhang2023radiomics}. However, they rarely analyze methodological dependencies, validation practices, or sources of variability systematically, leaving reproducibility and translational challenges insufficiently addressed.

This survey adopts an end-to-end, methodology-centric analysis of radiomics pipelines, independent of application domain, modeling paradigm, or imaging modality. It analyzes how interdependent design choices across acquisition, preprocessing, segmentation, feature extraction, feature selection, modeling, and evaluation jointly determine feature stability, validation reliability, and clinical translatability. By linking pipeline decisions to sources of variability and reproducibility failure, this review establishes a unifying analytical framework that complements existing application-driven, deep learning–centric, and modality-specific surveys.

The main contributions of this paper are summarized as follows:
\begin{itemize}
\item A unified, dependency-aware overview of radiomics pipelines from image acquisition and preprocessing to feature extraction and predictive modeling.
\item A comparative analysis of radiomic feature extraction, selection, dimensionality reduction, and modeling strategies, emphasizing their impact on robustness and reproducibility.
\item A validation-centric review of major clinical application domains, prioritizing evaluation design over reported performance metrics.
\item A critical discussion of open challenges and future research directions related to standardization, generalization, and clinical deployment.
\end{itemize}

The paper is organized as follows: Section 2 \label{sec2} presents radiomics pipeline. Section 3 \label{sec3} describes feature selection and dimensionality reduction. Section 4 \label{sec4} discusses radiomics modeling approaches. Section 5 \label{sec5} discusses the related work. Section 6 \label{sec6} presents evaluation protocols and validation. Section 7 \label{sec7} discusses challenges and limitations. Section 8 \label{sec8} discusses the future work. Section 9 \label{sec9} concludes the paper.

\section{Radiomics Pipeline}\label{sec2}
Radiomics follows a multi-stage pipeline that converts medical images into quantitative descriptors for predictive modeling. Each stage of this pipeline: (1) image acquisition; (2) segmentation; and (3) feature extraction and analysis, introduces methodological choices that directly influence feature stability, reproducibility, and downstream performance. It is important to understand these stages to interpret radiomics outcomes and evaluating their clinical reliability.

\begin{figure}[H]
    \centering
    \includegraphics[width=\linewidth]{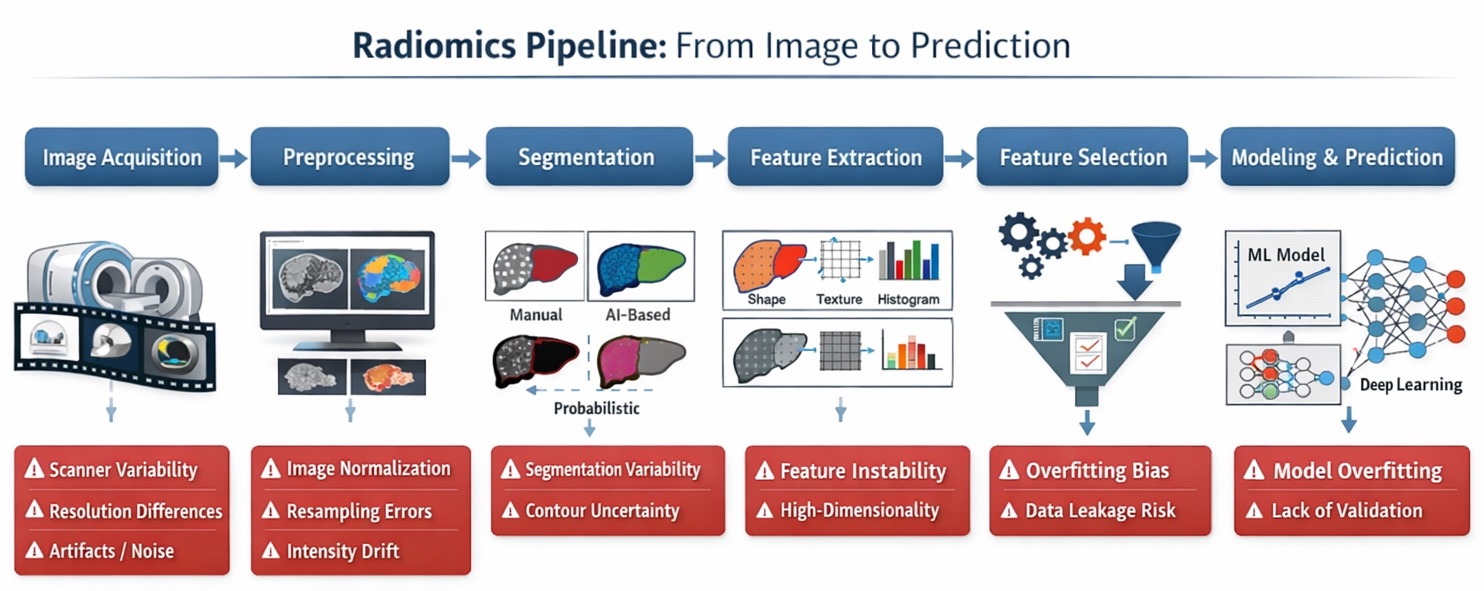}
    \caption{End-to-end radiomics pipeline with representative sources of variability and bias.}
    \label{fig:pipeline}
\end{figure}

Figure~\ref{fig:pipeline} provides an end-to-end radiomics pipeline illustrating the progression from image acquisition to predictive modeling, with representative sources of variability and bias at each stage, including acquisition heterogeneity, preprocessing effects, segmentation uncertainty, feature instability, data leakage, and model overfitting.

\subsection{Image Acquisition and Standardization}
Image acquisition is a primary source of variability in radiomics pipelines \cite{mi2020impact, zhao2021understanding}. Medical images are acquired using heterogeneous scanners and protocols, with variations in slice thickness, voxel spacing, reconstruction kernels, acquisition energy, and contrast timing. These factors directly influence image intensity distributions, spatial resolution, and noise characteristics, leading to systematic shifts in extracted radiomic features.

Inter-scanner variability and resolution differences impact texture and higher-order features, which are sensitive to interpolation, discretization, and noise statistics \cite{sang2024quantifying}. Inconsistent voxel anisotropy and resampling further impair feature stability and limit reproducibility across datasets, posing challenges for multi-center studies and external validation.

Standardization strategies aim to reduce acquisition-induced variability prior to feature extraction. Common approaches include:

\begin{enumerate}
    \item \textbf{Spatial normalization}, such as resampling to isotropic voxel spacing using linear or spline interpolation to reduce resolution-induced effects \cite{wang2022critical}.

    \item \textbf{Intensity normalization}, including z-score normalization, histogram matching, Nyúl’s piecewise linear mapping, and modality-specific scaling (e.g., Hounsfield unit windowing in CT) \cite{reinhold2019evaluating}.
    \item \textbf{Gray-level discretization}, using fixed bin-width or fixed bin-count schemes to stabilize texture matrix computation \cite{duron2019gray}.
    \item \textbf{Feature-level harmonization}, such as ComBat, Bayesian ComBat, and deep harmonization variants, to remove scanner-related batch effects while preserving biologically relevant variation \cite{hu2023image}.
\end{enumerate}

Acquisition variability affects radiomic feature reliability and stability, commonly quantified using intraclass correlation coefficients (ICC) \cite{xue2021radiomics}. Variations in scanner type, reconstruction kernel, and voxel resolution disproportionately reduce ICC values for texture and wavelet features, even under fixed segmentation. Shape features are comparatively robust, whereas higher-order features show the greatest sensitivity. Standardization and harmonization strategies mitigate these effects but introduce additional hyperparameters and modeling assumptions, requiring transparent reporting. 

Table~\ref{tab:acq_variability} summarizes key acquisition factors, affected feature categories, and mitigation strategies.

\begin{table}[!ht]
\caption{Acquisition-related sources of variability, affected radiomic feature categories, and commonly adopted mitigation strategies.}
\label{tab:acq_variability}
\centering
\begin{tabular}{lll}
\toprule
\textbf{Variability source} & \textbf{Affected features} & \textbf{Mitigation strategy} \\
\midrule
Slice thickness / resolution & Texture, wavelet & Isotropic resampling \\
Reconstruction kernel        & Texture, higher-order & Kernel harmonization \\
Scanner manufacturer         & Most feature categories & ComBat harmonization \\
Intensity scaling            & First-order, texture & Intensity normalization \\
\bottomrule
\end{tabular}
\end{table}

\subsection{Region of Interest (ROI) Segmentation}

Region of Interest (ROI) segmentation defines the spatial domain from which radiomic features are extracted and analyzed. Segmentation can be performed manually by expert annotators, semi-automatically using interactive algorithms, or fully automatically through model-based approaches.

Manual delineation remains common in clinical studies but it is time-consuming and prone to inter- and intra-observer variability \cite{joskowicz2019inter}. It is commonly quantified using overlap-based metrics such as the Dice similarity coefficient (DSC), defined as \begin{equation}
\mathrm{DSC} = \frac{2|A \cap B|}{|A| + |B|},
\label{eq:dice}
\end{equation}
where $A$ and $B$ denote two segmentations of the same anatomical structure.
Another commonly used overlap-based metric is the Jaccard Index (JI), also known as Intersection over Union (IoU), defined as
\begin{equation}
\mathrm{IoU} = \frac{|A \cap B|}{|A \cup B|} = \frac{|A \cap B|}{|A| + |B| - |A \cap B|}.
\label{eq:jaccard}
\end{equation}
The Jaccard Index provides a stricter measure of spatial agreement compared to DSC. The two metrics are monotonically related as
\begin{equation}
\mathrm{DSC} = \frac{2,\mathrm{IoU}}{1 + \mathrm{IoU}}, \qquad
\mathrm{IoU} = \frac{\mathrm{DSC}}{2 - \mathrm{DSC}}.
\end{equation}

Semi-automatic and automatic segmentation methods have been developed to address these limitations. Deep learning--based architectures such as U-Net and its variants, have become the common architectures for automated medical image segmentation across imaging modalities \cite{neha2025analytics}. These encoder--decoder models employ skip connections to preserve spatial resolution and are optimized using voxel-wise loss functions. This enables accurate delineation of complex anatomical structures when trained on sufficiently annotated datasets.

Despite methodological advances, segmentation uncertainty remains a major source of radiomics variability. Even minor ROI boundary perturbations, such as morphological dilation or erosion, can induce substantial fluctuations in shape descriptors and higher-order texture features \cite{zwanenburg2021image}. This sensitivity motivates robustness analysis and consensus-based strategies, including probabilistic segmentation and multi-observer aggregation, to improve reproducibility. 

\begin{figure}[t]
    \centering
    \includegraphics[width=\linewidth]{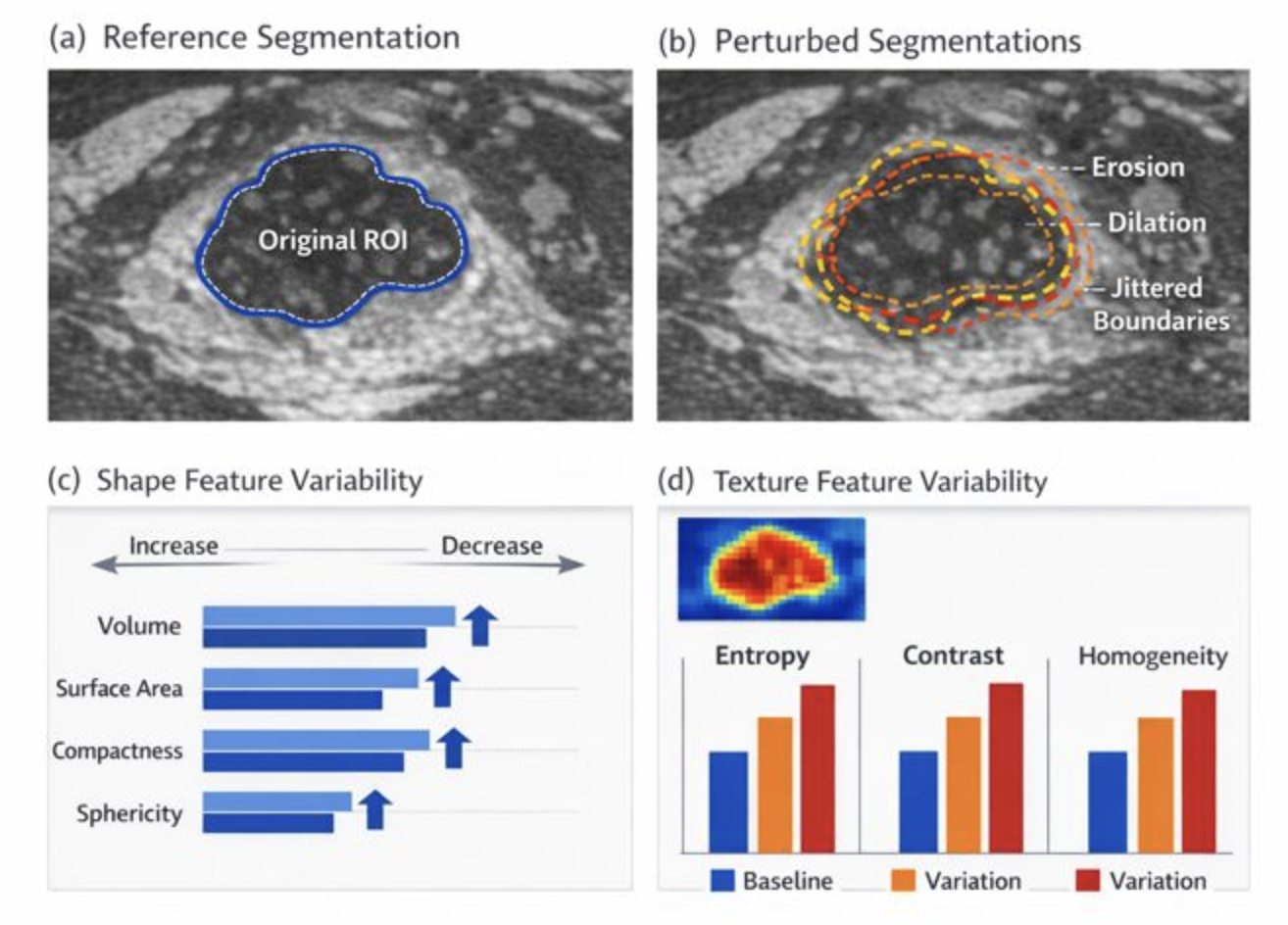}
    \caption{Effect of ROI boundary perturbations on shape and texture feature values, illustrating segmentation-induced variability. Fig. 2(a) shows an ROI. Fig. 2(b) represents boundary perturbations. Fig. 2(c) and Fig. 2(d) show their downstream effects on radiomic features.}
    \label{fig:roi_perturbation}
\end{figure}

Figure~\ref{fig:roi_perturbation} illustrates segmentation-induced variability. Shape features show systematic boundary-dependent changes, whereas texture features show amplified instability due to voxel reassignment at lesion margins.

\subsection{Radiomic Feature Extraction}

Radiomic feature extraction transforms segmented regions into quantitative descriptors that characterize underlying tissue phenotype \cite{vanGriethuysen2017radiomics}. These features are commonly grouped according to their mathematical formulation and the type of information they encode.

\begin{enumerate}
    \item \textbf{First-order statistics} describe the statistical distribution of voxel intensities within the ROI, capturing measures of central tendency, dispersion, and intensity range without accounting for spatial relationships. The representative features include the mean intensity 
    \begin{equation}
    \mu = \frac{1}{N} \sum_{i=1}^{N} x_i,
    \label{eq:mean}
    \end{equation}
    and variance
    \begin{equation}
    \sigma^2 = \frac{1}{N} \sum_{i=1}^{N} (x_i - \mu)^2,
    \label{eq:variance}
    \end{equation}
    where $x_i$ denotes voxel intensities and $N$ is the number of voxels.

    \item \textbf{Shape features} quantify geometric properties of the segmented region, including volume, surface area, compactness, elongation, and sphericity. These descriptors are invariant to intensity scaling and capture morphological characteristics relevant to disease phenotype.

    \item \textbf{Texture features} encode spatial relationships among voxel intensities and quantify intraregional heterogeneity. Common representations include gray-level co-occurrence matrices (GLCM), gray-level run-length matrices (GLRLM), gray-level size-zone matrices (GLSZM), neighboring gray-tone difference matrices (NGTDM), and gray-level dependence matrices (GLDM). For example, GLCM contrast is defined as
    \begin{equation}
    \sum_{i,j} (i - j)^2 P(i,j),
    \label{eq:glcm_contrast}
    \end{equation}
    where $P(i,j)$ denotes the joint probability of gray levels $i$ and $j$ at a specified spatial offset.
    
    \item \textbf{Higher-order features} are computed after applying image transformations such as wavelet decomposition, Laplacian-of-Gaussian filtering, Gabor filtering, or fractal analysis. These transformations enable multiscale and frequency-domain characterization by emphasizing structural patterns at different resolutions.

\end{enumerate}

Table~\ref{tab:feature_taxonomy} provides a taxonomy of radiomic feature categories, highlighting their mathematical basis, stability, interpretability, and sensitivity.

\begin{table}[!ht]
\centering
\caption{Taxonomy of radiomic feature categories and their key characteristics.}
\label{tab:feature_taxonomy}
\small
\begin{tabular}{p{1.5cm} p{2.9cm} p{1.6cm} p{2.2cm} p{1.6cm}}
\hline
\textbf{Feature category} & \textbf{Mathematical basis} & \textbf{Stability} & \textbf{Interpretability} & \textbf{Sensitivity} \\
\hline
First-order & Intensity statistics & High & High & Low \\
Shape & Geometric descriptors & High & High & Low \\
Texture & Spatial dependency matrices & Moderate & Moderate & High \\
Higher-order & Filtered or transformed features & Low--Moderate & Low & High \\
\hline
\end{tabular}
\end{table}

Radiomic feature extraction is commonly implemented using standardized software frameworks, including PyRadiomics \cite{vanGriethuysen2017radiomics}, Imaging Biomarker Explorer (IBEX) \cite{zhang2015ibex}, MaZda \cite{szczypinski2009mazda}, and the Computational Environment for Radiological Research (CERR) \cite{deasy2003cerr}. PyRadiomics is an open-source, Python-based library that provides Image Biomarker Standardisation Initiative -compliant feature extraction across first-order, shape, texture and higher order categories. IBEX is a MATLAB-based platform designed for quantitative imaging biomarker extraction, feature visualization, and sensitivity analysis in oncologic imaging. MaZda is a texture analysis software originally developed for medical image characterization, offering a broad range of statistical and model-based texture features. CERR is a MATLAB-based research framework developed for radiotherapy and imaging analysis, supporting radiomic feature extraction and multimodal data integration.

These tools support reproducibility but variations in parameter settings, discretization strategies, and preprocessing configurations remain a significant source of inter-study variability.

\section{Feature Selection and Dimensionality Reduction}
Radiomics yields high-dimensional and highly correlated feature spaces from limited cohorts, leading to instability and reduced generalization. Feature selection and dimensionality reduction are therefore essential for robust and reproducible modeling.

\subsection{Feature Selection Strategies}
Feature selection aims to identify a subset of informative and non-redundant features that preserve discriminative power and reduce model complexity. Common feature selection techniques are categorized as filter, wrapper, and embedded methods.

\begin{enumerate}
    \item \textbf{Filter methods} rank features independently of the predictive model using statistical relevance criteria. Representative techniques include variance thresholding, correlation analysis, and mutual information (MI) \cite{peng2005feature}, defined as
    \begin{equation}
    MI(X;Y) = \sum_{x,y} p(x,y)\log\frac{p(x,y)}{p(x)p(y)},
    \label{eq:mi}
    \end{equation}
    which quantifies statistical dependence between a feature $X$ and the target variable $Y$.

  A minimum redundancy--maximum relevance (mRMR) method extends this by prioritizing features that exhibit strong association with the outcome while reducing redundancy among selected features \cite{peng2005feature}. Filter methods offer computational efficiency and scalability in high-dimensional radiomics but they do not explicitly account for feature interactions or model-specific behavior.
  
  \item \textbf{Wrapper methods} integrate the learning algorithm into the selection process by iteratively evaluating feature subsets based on predictive performance \cite{kohavi1997wrappers}. Examples include recursive feature elimination (RFE), sequential forward selection, and genetic algorithms. These methods can capture complex feature interactions but they incur higher computational cost and exhibit increased susceptibility to overfitting in small-sample radiomics studies.
  
  \item \textbf{Embedded methods} perform feature selection during model training through regularization or intrinsic model constraints. Sparsity-inducing formulations such as LASSO \cite{tibshirani1996regression} and Elastic Net \cite{zou2005regularization} optimize the objective function. This results in compact and interpretable feature subsets.

 Embedded methods offer a balance between interpretability and predictive performance but their behavior remains influenced by feature scaling, correlation structure, and hyperparameter selection.
\end{enumerate}

Feature selection in radiomics shows instability under data resampling. Selected feature subsets vary substantially across cross-validation folds or cohort splits. Low selection frequency reflects limited robustness and weak generalization. Post-hoc feature importance analyses should be interpreted cautiously, as importance scores are highly sensitive to feature correlation, scaling, and sampling variability and do not imply stable or causal relevance.

\subsection{Dimensionality Reduction}

Dimensionality reduction complements explicit feature selection by projecting high-dimensional radiomic features into lower-dimensional latent representations.

Principal component analysis (PCA) is commonly employed to identify orthogonal components that capture maximal variance through eigen decomposition of the feature covariance matrix \cite{mackiewicz1993principal}. By retaining the leading components, PCA reduces redundancy and improves numerical stability, albeit at the expense of reduced feature interpretability.

Alternative approaches, including independent component analysis (ICA) \cite{lee1998independent}, partial least squares (PLS) \cite{cha1994partial}, and nonlinear manifold learning methods such as autoencoders \cite{bank2023autoencoders} and kernel PCA \cite{scholkopf1997kernel}, have been explored to capture higher-order dependencies and nonlinear structure in radiomics data. These methods increase representational compactness but introduce additional modeling complexity and stronger data requirements.

\begin{figure}[t]
    \centering
    \includegraphics[width=\linewidth]{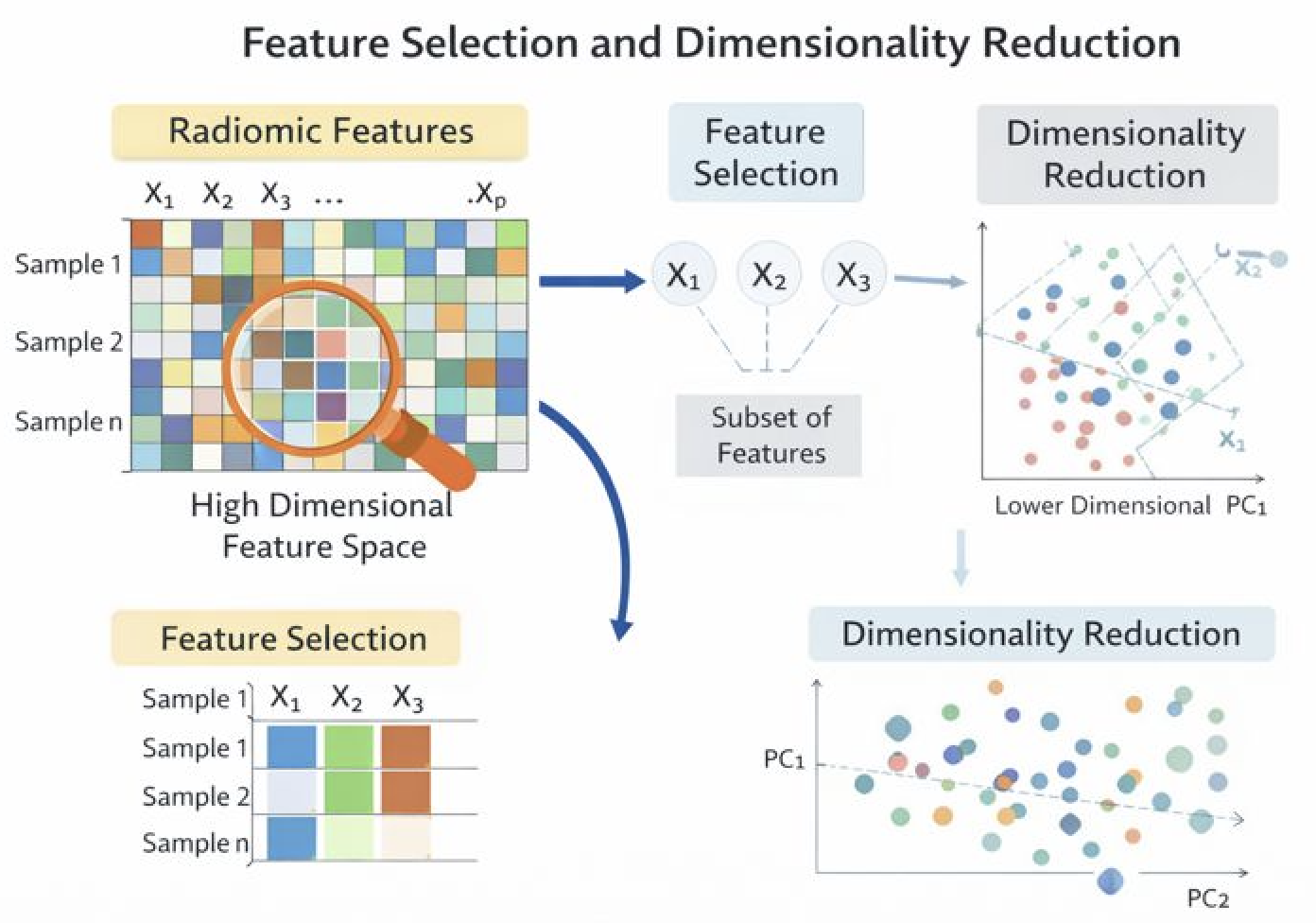}
    \caption{Schematic illustration of feature selection and dimensionality reduction in high-dimensional radiomic analysis.}
    \label{fig:FS}
\end{figure}

Figure~\ref{fig:FS}  illustrates the role of feature selection and dimensionality reduction in managing high-dimensional radiomic feature spaces. Feature selection identifies a compact subset of informative and non-redundant features from the original feature space, while dimensionality reduction projects the selected features into a lower-dimensional latent space, preserving dominant variance structures and improving modeling stability.

\subsection{Validation Considerations}
Feature selection and dimensionality reduction are closely coupled with the validation framework. Performing these steps outside the validation loop introduces information leakage and leads to optimistically biased performance estimates \cite{varma2006bias}. Embedding selection and reduction within cross-validation or nested validation schemes is therefore necessary for unbiased evaluation.

Leakage commonly arises when variance thresholding, mutual information ranking, LASSO regularization, or PCA are applied to the full dataset prior to data splitting. Such practices exploit test-set statistics during model construction, artificially inflating performance and yielding unstable feature subsets on independent evaluation. These effects are amplified in high-dimensional radiomics settings with limited sample sizes, directly undermining reproducibility, generalization, and translational reliability.

Table~\ref{tab:fs_comparison} summarizes common radiomics feature selection strategies and contrasts their robustness, leakage risk, and interpretability.

\begin{table}[!ht]
\centering
\caption{Feature selection strategies in radiomics and their methodological trade-offs.}
\label{tab:fs_comparison}
\begin{tabular}{llll}
\hline
\textbf{Method} & \textbf{Robustness} & \textbf{Leakage risk} & \textbf{Interpretability} \\
\hline
Filter & Low--Moderate & Moderate & High \\
Wrapper & Low & High & Moderate \\
Embedded & Moderate & Moderate & High \\
Dimensionality reduction & Moderate & Moderate & Low \\
\hline
\end{tabular}
\end{table}

\section{Radiomics Modeling Approaches}\label{sec4}

Radiomic features refined through selection or dimensionality reduction are incorporated into predictive models for clinical inference. Model choice directly affects accuracy, interpretability, robustness, and generalizability. Radiomics studies employ classical machine learning, deep learning–based, and hybrid modeling paradigms.

\subsection{Classical Machine Learning (CML)-Based Radiomics }

Classical machine learning (CML) algorithms remain widely adopted in radiomics due to their compatibility with handcrafted feature representations and effectiveness in limited-sample settings \cite{avanzo2020machine}. Commonly used models include support vector machines (SVM), logistic regression (LR), random forests (RF), and gradient-boosted (GB) trees (e.g., XGBoost and LightGBM). Additional approaches explored across applications include k-nearest neighbors (k-NN), naïve Bayes (NB) classifiers, decision trees (DT), and linear and quadratic discriminant analysis.

A key advantage of classical models lies in interpretability. Linear classifiers provide explicit decision functions and direct assessment of feature relevance. Tree-based ensembles offer feature importance measures that support qualitative model inspection. Data efficiency further contributes to their continued use in clinical radiomics studies.

At the same time, the performance of classical models remains tightly coupled to the stability and quality of handcrafted features. Sensitivity to feature redundancy, acquisition-induced variability, and preprocessing choices persists, while limited capacity to model complex, non-linear interactions constrains scalability as feature dimensionality increases.



\begin{figure}[H]
    \centering
    \includegraphics[width=\linewidth]{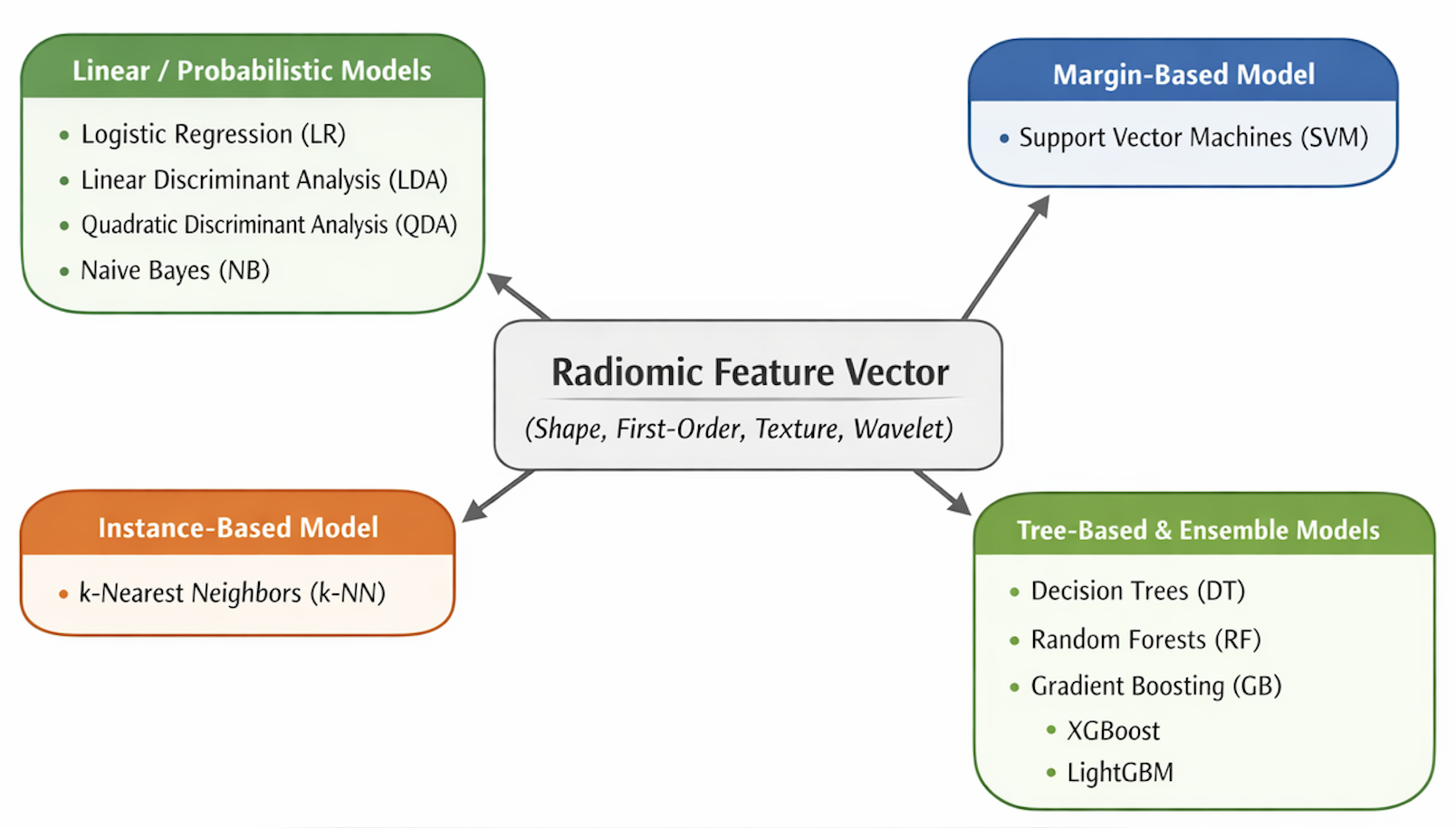}
    \caption{Classical Machine learning Models in Radiomics}
    \label{fig:ml}
\end{figure}

Figure \ref{fig:ml} Schematic overview of classical machine learning models commonly employed in radiomics, illustrating how handcrafted radiomic feature vectors (shape, first-order, texture, and wavelet features) are mapped to linear, margin-based, instance-based, and tree-based ensemble classifiers that are well suited for high-dimensional, limited-sample settings.

\subsection{Deep Learning (DL)--Based Radiomics}

Deep learning (DL) –based radiomics replaces handcrafted feature pipelines with end-to-end representation learning directly from imaging data \cite{avanzo2020machine}. Convolutional neural networks (CNNs) learn hierarchical, spatially structured features from raw or minimally processed images \cite{o2015introduction}. CNNs have shown strong performance in lesion detection, tumor classification, grading, and outcome prediction .

DL-based approaches enable modeling of complex imaging patterns that are difficult to capture using predefined descriptors and have shown promise across CT, MRI, PET, and digital pathology. However, their effectiveness is strongly dependent on data volume, annotation quality, and cohort diversity. Limited sample sizes and homogeneous datasets increase susceptibility to overfitting and restrict generalization. In addition, architectural variability, training stochasticity, and limited interpretability complicate reproducibility and clinical acceptance.

\subsection{Ensemble and Hybrid Radiomics Frameworks}

Ensemble and hybrid radiomics frameworks integrate multiple learners, feature spaces, or data modalities to mitigate limitations of individual models and improve robustness and generalization.

Boosting-based ensembles, including AdaBoost and GB, construct additive models by sequentially weighting weak learners and show strong performance when radiomic feature sets contain informative yet noisy descriptors \cite{parmar2015machine}. Sensitivity to feature instability and preprocessing variability, however, remains a limiting factor. Bagging-based approaches, such as RFs and random subspace methods, reduce variance through aggregation of decorrelated learners.

Stacking (stacked generalization) combines heterogeneous base learners—such as SVMs, RFs, and LR—via a meta-model trained on their predictive outputs \cite{naimi2018stacked}. Related formulations, including blending, Bayesian model averaging, and super learner frameworks, have also been explored to integrate complementary decision functions. In radiomics, stacking-based ensembles have shown improved robustness in heterogeneous datasets, provided strict separation between training and validation folds is maintained.

Fusion-based hybrid frameworks integrate radiomic features with DL representations, clinical variables, or multi-modal imaging data \cite{huang2016radiomics, ramachandram2017deep}. Feature-level fusion aggregates heterogeneous descriptors prior to modeling, whereas decision-level fusion combines predictions from independently trained models. Intermediate fusion strategies, including attention-based integration, graph-based modeling, and multi-task learning, have been analyzed to capture cross-modal dependencies.

Despite increased flexibility, ensemble and hybrid frameworks introduce additional complexity through multi-stage training and expanded hyperparameter spaces. This complexity challenges reproducibility, deployment, and clinical integration, underscoring the need for rigorous validation and transparent reporting.

\subsection{Radiomics vs.\ Deep Learning}\label{subsec:rad_vs_dl}

Radiomics and DL represent distinct and complementary approaches to quantitative medical image analysis. Radiomics relies on handcrafted, mathematically defined features combined with conventional ML models, whereas DL derives hierarchical feature representations directly from imaging data. These methodological differences have direct implications for data requirements, interpretability, robustness, validation complexity, and clinical translation.

Radiomics is well suited to limited-cohort studies and supports transparent, feature-level interpretation aligned with established imaging biomarkers. In contrast, DL excels in large-scale datasets and complex perceptual tasks but introduces challenges related to explainability, reproducibility, and deployment. Consequently, neither paradigm is universally optimal across clinical scenarios.

\begin{figure}[!ht]
    \centering
    \includegraphics[width=\linewidth]{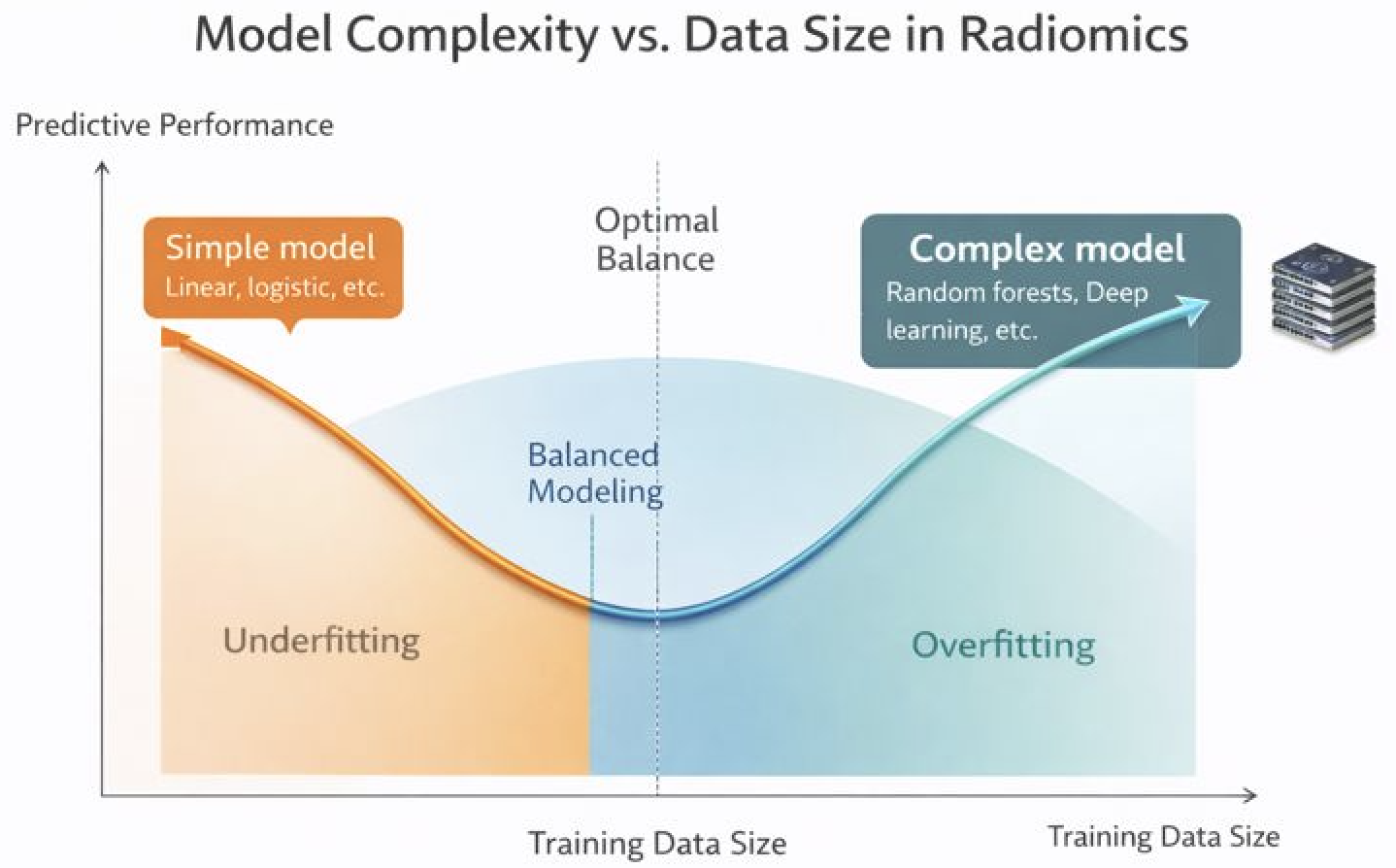}
    \caption{Model complexity vs. data size trade-off in radiomics}
    \label{fig:tradeoff}
\end{figure}

Figure~\ref{fig:tradeoff} illustrates the trade-off between model complexity and training data size in radiomics, highlighting underfitting, overfitting, and the setting for optimal generalization. Predictive performance depends on matching model capacity to available data, as simple models underfit high-dimensional feature spaces, whereas complex models overfit when data are limited.

Table~\ref{tab:rad_vs_dl} summarizes key methodological and practical distinctions between radiomics and DL approaches.

\begin{table}[t]
\centering
\caption{Comparison of radiomics and DL approaches in medical image analysis.}
\label{tab:rad_vs_dl}
\begin{tabular}{p{3.0cm} p{4.0cm} p{4.5cm}}
\hline
\textbf{Features} & \textbf{Radiomics} & \textbf{DL} \\
\hline
Data and annotation requirements & Effective with small cohorts; relies on accurate ROI delineation & Requires large, well-annotated datasets \\

Feature representation & Handcrafted, predefined, interpretable features & Automatically learned hierarchical representations \\

Interpretability & High feature-level transparency & Limited; relies on post-hoc explainability methods \\

Computational demands & Moderate; feasible on standard infrastructure & High; typically requires GPUs and extensive training \\

Robustness and generalization & Sensitive to acquisition variability; mitigated via harmonization & Sensitive to dataset bias; improves with data diversity and augmentation \\

Reproducibility and validation & Affected by preprocessing and feature stability; requires external validation & Affected by architectural and training stochasticity; requires large validation cohorts \\

Clinical translation & Higher acceptance and easier deployment & Slower adoption due to trust and infrastructure requirements \\

Representative use cases & Biomarker discovery, prognostic modeling, low-data studies & Detection, segmentation, large-scale prediction tasks \\
\hline
\end{tabular}
\end{table}

The growing adoption of hybrid radiomics frameworks reflects that radiomics and DL address complementary problems of clinical imaging analysis. Therefore, a comparative understanding of these frameworks is important for selecting appropriate modeling strategies and designing robust, clinically translatable studies.

\section{Related Work}\label{sec5}
Radiomics enables the extraction of quantitative imaging biomarkers that support disease characterization, prognostic assessment, and evaluation of therapeutic response. Features derived from baseline imaging have been associated with clinical outcomes, and longitudinal analysis captures temporal changes related to disease progression or treatment response. Across CT and MRI, radiomic analysis has been applied to nephrology, neurological, cardiovascular, pulmonary, and musculoskeletal conditions, facilitating objective assessment of structural and tissue-level alterations. 

Applications include outcome prediction in neurodegenerative disease and stroke, risk stratification in cardiovascular disorders using coronary CT angiography and cardiac MRI \cite{polidori2023radiomics}, characterization of parenchymal abnormalities in pulmonary disease, and quantitative evaluation of bone and cartilage degeneration in musculoskeletal imaging \cite{pleshkov2025radiomics}.

Yu et al. developed a CT-based radiomics and ML approach using histogram, texture, and gradient features combined with a linear SVM to differentiate renal tumor subtypes and oncocytoma. \cite{yu2017texture}. Lu et al. developed a multimodal MRI-based radiomics framework using intensity, texture, and shape features combined with hierarchical ML classifiers (primarily SVMs) to infer key molecular characteristics of gliomas and enable noninvasive stratification according to WHO-defined molecular subtypes \cite{lu2018machine}. Feng et al. proposed a SVM with RFE model, combined with a Synthetic Minority Oversampling Technique (SMOTE), for the quantitative texture-analysis of CT-images, to differentiate between different types of renal masses \cite{Feng2018MachineLearningCT}. SMOTE generates synthetic minority-class samples by interpolating between nearest-neighbor instances to address class imbalance during model training \cite{chawla2002smote}.

Chaddad et al. proposed a deep radiomics framework for survival prediction in recurrent glioblastoma by extracting CNN-based deep radiomic features from MRI and using RF to stratify patients into survival risk groups, showing the prognostic advantage of deep features over handcrafted descriptors \cite{chaddad2019deep}. Li et al. developed a deep CNN–based survival prediction framework for rectal cancer using PET/CT imaging, incorporating spatial pyramid pooling to accommodate variable tumor sizes \cite{li2019deep}. Chen et al. developed a radiomics-based ML framework using texture features extracted from MRI and multiple feature-selection–classifier combinations, showing that distance correlation–based feature selection combined with linear discriminant analysis (LDA) or LR can differentiate glioblastoma from metastatic brain tumors \cite{chen2019radiomics}. 

Yi et al. applied SVM and RFs, with radiomic features, to differentiate between low-grade and high-grade clear cell renal cell carcinoma (ccRCC) – a subtype of RCC \cite{Yi2020CTRadiomicsRCC}. Gitto et al. developed a radiomics–based ML model using first-order and texture features extracted from MRI and an AdaBoost ensemble classifier to differentiate low-grade from high-grade cartilaginous bone tumors \cite{gitto2020mri}. Deng et al. applied CT texture analysis using a filtration–histogram method with various spatial scaling filters to derive features which capture heterogeneity \cite{Deng2020CTTextureRenal}. Various statistical metrics such as Entropy, kurtosis, skewness, and mean positive pixel-features were extracted from CT images, with ROI drawn manually on the largest tumor cross-section. The model was assessed using LR analysis to segregate benign and malignant tumors. Erdim et al. employed DT, k-NN, LR, SVM, NB, RF, Feed Forward Neural Network (FFN), and locally weighted learning, to separate benign from malignant renal tumors using texture features of CT scans \cite{erdim2020prediction}. Their feature-selection strategy utilizes greedy-search for optimizing feature-set. Pie et al. proposed statistical analysis of a radiomics nomogram that incorporates a radiomics signature and clinical factors for the preoperative differentiation between fat deposits and ccRCC \cite{nie2020ct}. 

Sun et al. proposed a SVM based approach, combining qualitative radiologic features with quantitative texture features, to differentiate benign from malignant renal tumors using CT-images \cite{sun2020radiologic}. Uhlig et al. developed a standalone XGBoost classifier and LR to classify five renal tumor subtypes \cite{uhlig2020discriminating}. Both techniques use RFE to prune undesired features.

Wang et al. applied RF, SVM, and LR from CT images to differentiate ccRCC from non-ccRCC \cite{Wang2021RadiomicsCT_RCC}. Correlation analysis removed redundant features. LR identified key predictors: variance, High Gray Level Run Emphasis (HGLRE) and minimum intensity. Gurbani et al. evaluated a CT-based radiomics and ML framework for identifying aggressive tumor features in RCC, focusing on high nuclear grade and sarcomatoid differentiation in large RCCs using non-contrast and portal venous phase CT images \cite{gurbani2021evaluation}. Volumetric radiomic features were extracted from 3D tumor segmentations, and multiple ML classifiers (XGBoost, RF, and SVM), were investigated with feature ranking and selection strategies.

Alhussaini et al. applied RF, SVM, KNN, LR, and NB to differentiate malignant tumors using limited handcrafted radiomics features extracted from CT scans \cite{alhussaini2022comparative}. Features are reduced using sparsity-driven regularization that eliminates less informative variables. The filtered features are classified using ML techniques. Lam et al. developed an MRI-based radiomics model using handcrafted features and a LightGBM classifier to predict tumor mutational burden in lower-grade gliomas \cite{lam2022radiomics}. 

He et al. proposed an ensemble framework for malignancy risk-prediction in cystic renal lesions using CT-scans \cite{He2023DeepRadiomicsCysticRenal}. The method integrates handcrafted radiomics features with DL-features extracted from a pretrained residual network. GB, XGBoost and DT are employed to combine the derived features for the classification. Kumar et al. evaluated MRI radiomic features with five ML classifiers (SVM, RF, GB, NB, and AdaBoost) for low- vs. high-grade glioma classification \cite{kumar2023machine}. Xu et al. integrated radiomics on CT-scans with clinical attributes (demographics, vital signs, and comorbidities), ML (RF and XGBoost) and CNN for binary classification of renal tumors \cite{xu2023classification}.

Magnuska et al. integrated radiomics and DL-based features for ultrasound (US) based binary breast-tumor classification \cite{Magnuska2024Radiol232554}. The model employs SVM, RF and LR for tumor categorization.

Chaddad et al. developed an MRI-based radiomics framework using XGBoost and RF with feature selection to differentiate lower-grade gliomas from glioblastoma and to derive radiomic risk signatures from pre-treatment scans for tumor classification and survival prediction \cite{chaddad2025radiomic}. Kilicarslan et al. proposed an ensemble deep learning framework for RCC subtype classification using MRI data \cite{kilicarslan2025fusion}. Features are extracted via transfer learning using pretrained DenseNet architectures, followed by Global Average Pooling (GAP) to aggregate spatial activations into compact representations. The resulting feature vectors are concatenated and classified using SVM.

Table \ref{tab:rw_part1} and Table \ref{tab:rw_part2} presents a comparison of classical radiomics and ML-based studies.

\begin{table*}[!ht]
\centering
\caption{Classical radiomics-based machine learning studies for tumor characterization and classification}
\label{tab:rw_part1}
\begin{tabular}{p{2.0cm} p{3.0cm} p{3.3cm} p{3.3cm}}
\hline
\textbf{Study (Year)} & \textbf{Modality and clinical task} & \textbf{Feature and model} & \textbf{Key limitations} \\
\hline
Yu et al. (2017) & CT; differentiation of renal tumor subtypes and oncocytoma & Histogram, texture, gradient features; linear SVM & Handcrafted features only; limited robustness to segmentation variability; single classifier \\

Lu et al. (2018) & MRI; glioma molecular subtype stratification & Intensity, texture, shape features; ML classifiers (SVM) & High-dimensional handcrafted features; limited generalization across scanners and protocols \\

Feng et al. (2018) & CT; differentiation of renal mass types & Texture features; SVM with RFE and SMOTE & Synthetic oversampling introduces bias; manual ROI delineation \\

Chaddad et al. (2019) & MRI; survival prediction in recurrent glioblastoma & CNN-based deep radiomic features; RF & Limited interpretability; data-intensive deep features \\

Li et al. (2019) & PET/CT; survival prediction in rectal cancer & Deep CNN with spatial pyramid pooling & Requires large datasets; limited explainability \\

Chen et al. (2019) & MRI; glioblastoma vs. metastatic brain tumors & Texture features; distance correlation feature selection with LDA/LR & Sensitivity to feature selection strategy; handcrafted features \\

Yi et al. (2020) & CT; grading of clear cell RCC (low vs. high grade) & Radiomic features; SVM and RF & Binary grading task only; limited subtype coverage \\
\hline
\end{tabular}
\end{table*}

\begin{table*}[!ht]
\centering
\caption{Classical radiomics-based machine learning studies for tumor characterization and classification}
\label{tab:rw_part2}
\begin{tabular}{p{2.0cm} p{3.0cm} p{3.3cm} p{3.3cm}}
\hline
\textbf{Study (Year)} & \textbf{Modality and clinical task} & \textbf{Feature and model} & \textbf{Key limitations} \\
\hline
Gitto et al. (2020) & MRI; grading of cartilaginous bone tumors & First-order and texture features; AdaBoost & Limited evaluation across tumor types; potential overfitting \\

Deng et al. (2020) & CT; benign vs. malignant renal tumor differentiation & Filtration–histogram texture features; LR & ROI drawn on single slice; limited 3D tumor representation \\

Erdim et al. (2020) & CT; benign vs. malignant renal tumors & Texture features; DT, k-NN, LR, SVM, NB, RF, FFN & Extensive model comparison without unified optimization; handcrafted features \\

Nie et al. (2020) & CT; fat-poor angiomyolipoma vs. ccRCC & Radiomics signature with clinical factors; nomogram & Limited external validation; dependence on clinical variables \\

Sun et al. (2020) & CT; benign vs. malignant renal tumors & Qualitative radiologic + quantitative texture features; SVM & Manual feature design; limited scalability \\

Uhlig et al. (2020) & CT; five renal tumor subtype classification & Radiomic features; XGBoost and LR with RFE & Feature pruning sensitive to training data; class imbalance \\

Wang et al. (2021) & CT; ccRCC vs. non-ccRCC classification & Radiomic features; RF, SVM, LR & Limited interpretability of ensemble models; correlation-based feature removal \\

Gurbani et al. (2021) & CT (non-contrast and portal venous); aggressive RCC phenotype prediction & 3D volumetric radiomic features; XGBoost, RF, SVM & Focus on large tumors only; complex feature selection pipeline \\
\hline
\end{tabular}
\end{table*}

Table \ref{tab:rw_part3} presents a comparison of deep radiomics and hybrid radiomics–DL frameworks.

\begin{table*}[!ht]
\centering
\caption{A comparison of deep radiomics and hybrid radiomics–DL frameworks}
\label{tab:rw_part3}
\begin{tabular}{p{2.3cm} p{3.5cm} p{3.3cm} p{3.4cm}}
\hline
\textbf{Study (Year)} & \textbf{Modality and clinical task} & \textbf{Feature and model} & \textbf{Key limitations} \\
\hline
Alhussaini et al. (2022) & CT; malignant renal tumor differentiation & Handcrafted radiomic features; sparsity-driven feature reduction with RF, SVM, k-NN, LR, NB & Limited feature diversity; reliance on handcrafted descriptors \\

He et al. (2023) & CT; malignancy risk prediction in cystic renal lesions & Handcrafted radiomics + DL features from pretrained ResNet; GB, XGBoost, DT & Increased model complexity; feature fusion strategy not fully interpretable \\

Xu et al. (2023) & CT; binary renal tumor classification & Radiomics + clinical attributes with RF, XGBoost, and CNN & Heterogeneous data integration; potential clinical data dependency \\

Kilicarslan et al. (2025) & MRI; RCC subtype classification & Transfer learning with pretrained DenseNet, GAP-based deep features; SVM & Dependence on pretrained models; limited interpretability of deep features \\
Lam et al. (2022) & MRI; tumor mutational burden prediction in lower-grade gliomas & Handcrafted radiomic features; LightGBM & Limited biological interpretability; scanner variability sensitivity \\

Kumar et al. (2023) & MRI; low- vs. high-grade glioma classification & Radiomic features; SVM, RF, GB, NB, AdaBoost & Handcrafted features only; binary grading task \\

Magnuska et al. (2024) & Ultrasound; binary breast tumor classification & Radiomics + DL features; SVM, RF, LR & Operator-dependent US acquisition; limited generalization \\

Chaddad et al. (2025) & MRI; glioma grading and survival prediction & Radiomic features with selection; XGBoost, RF & Feature-selection sensitivity; survival modeling complexity \\
\hline
\end{tabular}
\end{table*}

\section{Evaluation Protocols and Validation}\label{sec6}
Evaluation protocols determine the reliability, generalization, and translational potential of radiomics models. Given the high dimensionality of radiomic feature spaces and the limited size of most imaging cohorts, rigorous validation is required to avoid biased performance estimates. Robust evaluation requires strict control of train/test separation, principled resampling strategies, external validation, and comprehensive performance assessment.

\subsection{Train/Test Leakage and Cross-Validation}
Train/test leakage occurs when information from evaluation data influences model development. Data leakage commonly arises when preprocessing steps such as normalization, harmonization, feature selection, or dimensionality reduction are performed before cross-validation \cite{sasse2025overview}. In radiomics, the high feature-to-sample ratio amplifies the impact of such leakage, leading to inflated performance estimates.

Let $D = \{(x_i, y_i)\}_{i=1}^{N}$ denote a dataset. In $k$-fold cross-validation, $D$ is partitioned into disjoint subsets $\{D_1, \ldots, D_k\}$. For each fold $j$, a model is trained on $D \setminus D_j$ and evaluated on $D_j$. The estimated performance metric $\hat{M}$ is
\begin{equation}
\hat{M} = \frac{1}{k} \sum_{j=1}^{k} M(D_j),
\end{equation}
where $M(\cdot)$ denotes the evaluation metric.

Hyperparameter tuning, feature selection, and dimensionality reduction must be restricted to the training portion of each fold. Nested cross-validation enforces this constraint by introducing an inner loop for model selection and an outer loop for unbiased performance estimation.

Bootstrapping is employed to estimate performance variability. Given $B$ bootstrap samples drawn with replacement from $D$, confidence intervals can be computed from the empirical distribution of the performance metric.

Given $B$ bootstrap resamples, the $(1-\alpha)$ confidence interval for a performance metric $M$ is estimated as
\begin{equation}
\text{CI}_{1-\alpha} = \left[ M^{(\alpha/2)}, \; M^{(1-\alpha/2)} \right],
\end{equation}
where $M^{(q)}$ denotes the $q$-quantile of the empirical bootstrap distribution of $M$.

\subsection{External and Temporal Validation}
External validation evaluates a trained model on an independent dataset collected under different acquisition conditions, scanners, or institutions. This setting tests robustness to distributional shift, which is common in radiomics due to scanner heterogeneity and protocol variability.

Temporal validation represents a related strategy in which models are trained on earlier cases and evaluated on data acquired at a later time point. This approach assesses stability under evolving clinical practice and acquisition settings.

Despite their importance, external and temporal validation remain underrepresented in radiomics studies. Reported performance frequently declines under independent testing, indicating sensitivity to cohort composition and acquisition variability.

\begin{figure}[t]
    \centering
    \includegraphics[width=\linewidth]{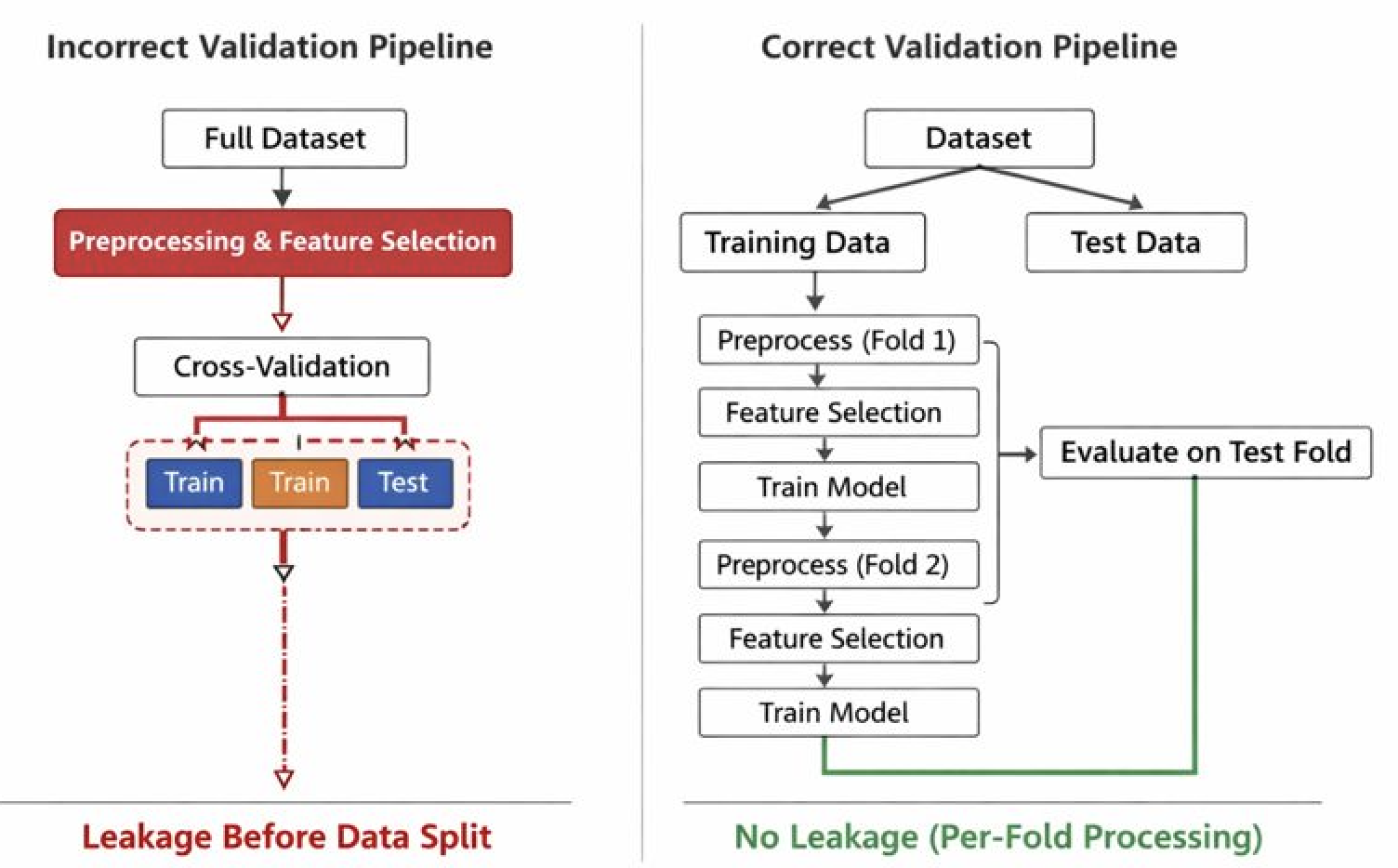}
    \caption{Comparison of incorrect and correct radiomics validation pipelines.}
    \label{fig:validation}
\end{figure}

Figure \ref{fig:validation} illustrates incorrect versus correct validation pipelines in radiomics. The left panel illustrates a common data leakage scenario in which preprocessing and feature selection are performed on the full dataset prior to cross-validation, resulting in optimistically biased performance estimates. The right panel depicts a leakage-free evaluation protocol, where data splitting precedes all preprocessing and feature selection steps, which are executed independently within each training fold before evaluation on held-out data.

\subsection{Performance Metrics}
Accuracy is insufficient for evaluating radiomics models, particularly in imbalanced datasets. Discriminative ability is commonly assessed using the area under the receiver operating characteristic curve (AUC), which evaluates ranking performance independent of decision thresholds \cite{richardson2024receiver}.

For binary classification: sensitivity, specificity, precision and F1-score are reported to account for class imbalance.



Calibration assesses agreement between predicted probabilities $\hat{p}_i$ and observed outcomes $y_i$ \cite{huang2020tutorial}. Brier score provides a quantitative measure of probabilistic accuracy \cite{rufibach2010use}.
\begin{equation}
\text{Brier} = \frac{1}{N} \sum_{i=1}^{N} (\hat{p}_i - y_i)^2.
\end{equation}
Calibration curves and goodness-of-fit tests further characterize systematic deviations between predicted and observed risks. Statistical comparison of AUCs can be performed using DeLong’s test in paired evaluation settings.

\subsection{Stability and Statistical Significance Analysis}

Feature and model stability are critical considerations in radiomics. Stability is commonly assessed using test--retest analysis or perturbation-based resampling and quantified via the intraclass correlation coefficient (ICC):

\begin{equation}
\text{ICC} = \frac{\sigma^2_{\text{between}}}{\sigma^2_{\text{between}} + \sigma^2_{\text{within}}}.
\end{equation}

Features or models with low ICC values reflect sensitivity to acquisition or segmentation variability and should be excluded to improve reproducibility \cite{xue2021radiomics}. Permutation testing is used to assess the statistical significance of model performance by comparing observed results against a null distribution generated through random label permutations, thereby guarding against spurious associations in high-dimensional feature spaces.


\section{Challenges and Limitations}\label{sec7}
\begin{figure}[!ht]
    \centering
    \includegraphics[width=\linewidth]{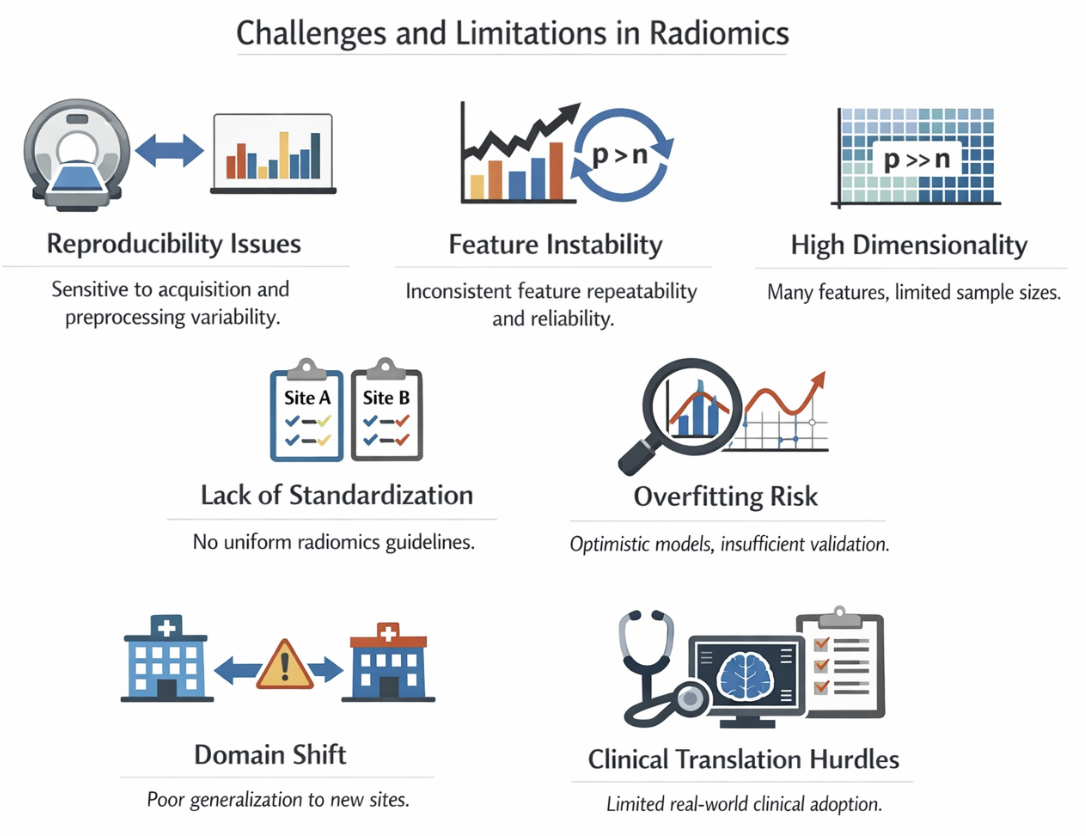}
    \caption{Overview of key challenges and limitations in radiomics.}
    \label{fig:limitation}
\end{figure}

Despite sustained methodological progress, radiomics continues to face fundamental limitations that impede reproducibility, generalization, and clinical translation as shown in \ref{fig:limitation}. These challenges arise from the sensitivity of handcrafted features to acquisition and preprocessing variability, intrinsic statistical constraints of high-dimensional modeling under limited data, and persistent weaknesses in validation and standardization practices.

\subsection{Reproducibility and Pipeline Sensitivity}
Radiomic features show strong dependence on imaging acquisition parameters, reconstruction algorithms, preprocessing choices, and segmentation strategies. Even minor variations in scanner settings, voxel resolution, intensity discretization, or region-of-interest delineation can substantially alter feature distributions, with higher-order texture features being particularly affected. This sensitivity produces pronounced inter-site variability and undermines robustness under domain shift, leading to degraded performance when models trained on single-center data are evaluated on external cohorts.

\subsection{Feature Stability and Reliability}
Radiomic features present heterogeneous stability under test--retest conditions and controlled perturbations. First-order intensity statistics and shape descriptors generally show higher repeatability, whereas texture- and filter-based features are highly sensitive to noise, quantization, and spatial resolution. Unstable features compromise biomarker interpretability and model reliability, particularly when feature selection prioritizes discriminative power without accounting for robustness. Inconsistent reporting of stability analyses further limits reproducibility and cross-study comparison.

\subsection{High Dimensionality and Limited Sample Sizes}
Radiomics typically operates in a high-dimensional settings ($p \gg n$), where hundreds to thousands of features are extracted from relatively small patient cohorts. This imbalance results in variance inflation, unstable parameter estimates, and increased risk of spurious associations. Although feature selection and regularization mitigate dimensionality, they do not overcome fundamental constraints related to identifiability, statistical power, and uncertainty estimation. These issues are amplified in multi-class classification and survival analysis, where reliable stratified validation and subgroup analysis are often infeasible.

\subsection{Lack of Standardization Across Studies}
The absence of end-to-end standardization across radiomics pipelines remains a major barrier to reproducibility. Variability in acquisition protocols, preprocessing workflows, feature definitions, discretization schemes, software implementations, and validation designs leads to inconsistent feature representations and non-comparable results. While standardized feature sets and reporting guidelines have been proposed, their adoption remains uneven, limiting cumulative evidence synthesis and robust meta-analytic evaluation.

\subsection{Overfitting and Validation Bias}
Overfitting is pervasive in radiomics due to high feature dimensionality, extensive model and hyperparameter exploration, and insufficiently rigorous validation strategies. Feature selection and tuning performed outside nested validation frameworks introduce optimistic bias, inflating reported performance. Increased model complexity through ensemble or hybrid approaches further exacerbate this issue under data-limited conditions. Performance metrics are often reported without adequate uncertainty quantification, obscuring true generalization capability.

\subsection{Domain Shift and Generalization Failure}
Radiomics models are particularly vulnerable to domain shift arising from differences in imaging protocols, patient populations, and clinical practice patterns. Harmonization methods reduce batch effects but are insufficient to address complex, non-linear shifts in feature distributions. Consequently, models demonstrating strong retrospective performance fail when deployed in heterogeneous real-world clinical settings.

\subsection{Barriers to Clinical Translation}
Despite promising retrospective findings, clinical adoption of radiomics remains limited. Most studies lack prospective or longitudinal validation and provide insufficient evidence of clinical utility. Translation is further hindered by complex preprocessing requirements, dependence on accurate segmentation, computational overhead, limited interpretability, and inadequate uncertainty modeling. Regulatory considerations and the absence of standardized deployment frameworks continue to restrict clinician trust and real-world integration.
Overall, radiomics pipelines remain fragile to acquisition and preprocessing variability, statistically constrained by high-dimensional modeling under limited data, and predominantly evaluated using retrospective designs. While methodological advances have increased feature complexity and model capacity, they have not resolved core issues related to stability, standardization, and generalization, resulting in performance degradation under independent validation and clinical deployment.

\section{Discussion and Future Work}\label{sec8}
Radiomics has shown sustained relevance in quantitative medical image analysis, however its future impact depends on methodological refinement, integration with emerging learning models, and alignment with clinical practice. This section outlines key directions that build on established strengths while addressing persistent limitations.

\subsection{Interpretability and Feature Relevance}
Radiomics offers explicit feature definitions, enabling feature-level interpretability not inherently available in end-to-end DL models. Feature relevance can be examined through coefficient analysis, permutation importance, and stability metrics under resampling or perturbation. These analyses support traceability of model decisions to specific image-derived properties.

However, statistical relevance derived from model optimization does not guarantee clinical relevance. Feature importance rankings are sensitive to correlation structure, regularization strength, and sampling variability. Future radiomics studies should distinguish predictive contribution from feature robustness by jointly reporting effect size, selection frequency across resampling, and test–retest stability. Feature relevance analysis that ignores stability constraints risks promoting non-reproducible biomarkers.

\subsection{Hybrid Radiomics–DL and Transformer-Based Models}
Hybrid frameworks combine handcrafted radiomic features with learned representations to extract complementary information. DL components operate either on image patches or feature embeddings, while radiomics provides structured, low-dimensional descriptors. Transformer-based architectures extend this framework by modeling feature interactions through self-attention mechanisms rather than fixed convolutional locality.

In hybrid radiomics–transformer models, radiomic features can be treated as tokens, enabling attention-based weighting and interaction modeling. This allows adaptive feature relevance estimation and long-range dependency modeling. Technical challenges include feature scaling compatibility, attention collapse in low-samples, and increased variance due to model capacity. These models require strict regularization and external validation to avoid capacity-driven overfitting.

\subsection{Multimodal Fusion}
Radiomics supports multimodal fusion through its compatibility with heterogeneous data types. Fusion strategies can be categorized as feature-level, intermediate, or decision-level. Feature-level fusion concatenates modality-specific representations, whereas intermediate fusion aligns latent spaces through joint embedding or attention mechanisms. Decision-level fusion aggregates independent predictions using weighted or probabilistic schemes.

Future radiomics research should favor fusion strategies that explicitly model modality uncertainty and conditional dependence. Naive concatenation amplifies noise and correlation effects. Modality-aware weighting, attention-based fusion, and Bayesian integration offer more principled alternatives, particularly in settings with missing or partially observed modalities.

\subsection{Self-Supervised and Representation Learning}
Self-supervised learning provides a mechanism for representation learning without manual labels. Common objectives include contrastive learning, reconstruction-based learning, and predictive pretext tasks. In radiomics, self-supervised pretraining can be applied at the image or region level to improve feature robustness prior to downstream modeling.

Integration of self-supervised representations with handcrafted radiomic features raises several technical questions, including representation alignment, redundancy control, and interpretability preservation. Empirical evaluation should assess whether self-supervised features improve generalization under domain shift and whether they maintain stability under acquisition variability.

\subsection{Federated Radiomics}
Federated learning enables distributed model training across institutions without centralized data aggregation. In radiomics, federated settings introduce non-identically distributed data due to scanner, protocol, and population differences. These factors complicate optimization and convergence.

Technical challenges include client drift, communication efficiency, and aggregation bias. Methods such as weighted aggregation, domain-aware optimization, and federated feature normalization require further investigation. Evaluation of federated radiomics models should include cross-site generalization and stability analysis rather than aggregated performance alone.

\subsection{Standard Benchmarks and Reporting Standards}

Radiomics lacks standardized benchmarks that support reproducible method comparison. Existing studies vary widely in task definition, cohort composition, preprocessing, and validation design. This heterogeneity limits cross-study comparison and cumulative evidence synthesis.

Future benchmarks should define fixed training–validation–test splits, standardized preprocessing pipelines, and reference evaluation metrics. Reporting standards should mandate disclosure of feature definitions, discretization parameters, validation nesting, and uncertainty estimates. Without such standardization, performance comparisons remain inconclusive and clinically uninformative.

\section{Conclusion}\label{sec9}
Radiomics provides a structured framework for quantitative medical image analysis, offering interpretable feature representations that remain effective in limited-data clinical settings. This survey presented an end-to-end, methodology-centric analysis of radiomics pipelines, emphasizing how design choices across acquisition, preprocessing, segmentation, feature engineering, modeling, and validation jointly determine reproducibility, robustness, and translational validity.

Despite sustained progress, radiomics is constrained by feature instability, sensitivity to pipeline variability, high dimensionality under limited sample sizes, and persistent validation bias. Inconsistent standardization and limited external evaluation further restrict generalization and clinical adoption. These limitations underscore that reported performance gains often reflect methodological artifacts rather than robust predictive capability.

This review is limited by its focus on methodological analysis rather than quantitative performance aggregation, as heterogeneity across datasets, tasks, and validation protocols precludes meaningful meta-analysis. Coverage reflects representative trends in the literature rather than exhaustive benchmarking or disease-specific optimization.

Future progress depends on standardized pipelines, stability-aware feature selection, leakage-free validation, and principled integration with DL and multimodal frameworks. Radiomics remains a valuable but fragile paradigm; its long-term impact will be determined by methodological rigor rather than increasing model complexity alone.

\section*{Declarations}

\subsection*{Ethics, Consent to Participate, and Consent to Publish}
Not applicable.

\subsection*{Competing interests}
The author declares that there are no competing interests.

\subsection*{Author contributions}
All authors whose names appear on the submission contributed equally to this work. 

\subsection*{Availability of data and materials}
Not applicable. This manuscript does not report data generation or analysis.

\bibliography{sn-bibliography}

@article{neha2025analytics,
  title={An analytics-driven review of U-Net for medical image segmentation},
  author={Neha, Fnu and Bhati, Deepshikha and Shukla, Deepak Kumar and Dalvi, Sonavi Makarand and Mantzou, Nikolaos and Shubbar, Safa},
  journal={Healthcare Analytics},
  pages={100416},
  year={2025},
  publisher={Elsevier}
}

@article{huang2016radiomics,
  title={Radiomics signature: a potential biomarker for the prediction of disease-free survival in early-stage (I or II) non—small cell lung cancer},
  author={Huang, Yanqi and Liu, Zaiyi and He, Lan and Chen, Xin and Pan, Dan and Ma, Zelan and Liang, Cuishan and Tian, Jie and Liang, Changhong},
  journal={Radiology},
  volume={281},
  number={3},
  pages={947--957},
  year={2016},
  publisher={Radiological Society of North America}
}

@article{rufibach2010use,
  title={Use of Brier score to assess binary predictions},
  author={Rufibach, Kaspar},
  journal={Journal of clinical epidemiology},
  volume={63},
  number={8},
  pages={938--939},
  year={2010},
  publisher={Elsevier}
}

@article{huang2020tutorial,
  title={A tutorial on calibration measurements and calibration models for clinical prediction models},
  author={Huang, Yingxiang and Li, Wentao and Macheret, Fima and Gabriel, Rodney A and Ohno-Machado, Lucila},
  journal={Journal of the American Medical Informatics Association},
  volume={27},
  number={4},
  pages={621--633},
  year={2020},
  publisher={Oxford University Press}
}

@article{richardson2024receiver,
  title={The receiver operating characteristic curve accurately assesses imbalanced datasets},
  author={Richardson, Eve and Trevizani, Raphael and Greenbaum, Jason A and Carter, Hannah and Nielsen, Morten and Peters, Bjoern},
  journal={Patterns},
  volume={5},
  number={6},
  year={2024},
  publisher={Elsevier}
}

@article{sasse2025overview,
  title={Overview of leakage scenarios in supervised machine learning},
  author={Sasse, L and Nicolaisen-Sobesky, E and Dukart, J and Eickhoff, SB and G{\"o}tz, M and Hamdan, S and Komeyer, V and Kulkarni, A and Lahnakoski, JM and Love, Bradley Carl and others},
  journal={Journal of Big Data},
  volume={12},
  number={1},
  pages={135},
  year={2025},
  publisher={Springer}
}

@article{chawla2002smote,
  title={SMOTE: synthetic minority over-sampling technique},
  author={Chawla, Nitesh V and Bowyer, Kevin W and Hall, Lawrence O and Kegelmeyer, W Philip},
  journal={Journal of artificial intelligence research},
  volume={16},
  pages={321--357},
  year={2002}
}

@article{ramachandram2017deep,
  title={Deep multimodal learning: A survey on recent advances and trends},
  author={Ramachandram, Dhanesh and Taylor, Graham W},
  journal={IEEE signal processing magazine},
  volume={34},
  number={6},
  pages={96--108},
  year={2017},
  publisher={IEEE}
}

@article{parmar2015machine,
  title={Machine learning methods for quantitative radiomic biomarkers},
  author={Parmar, Chintan and Grossmann, Patrick and Bussink, Johan and Lambin, Philippe and Aerts, Hugo JWL},
  journal={Scientific reports},
  volume={5},
  number={1},
  pages={13087},
  year={2015},
  publisher={Nature Publishing Group UK London}
}

@article{naimi2018stacked,
  title={Stacked generalization: an introduction to super learning},
  author={Naimi, Ashley I and Balzer, Laura B},
  journal={European journal of epidemiology},
  volume={33},
  number={5},
  pages={459--464},
  year={2018},
  publisher={Springer}
}

@article{zwanenburg2021image,
  title={The image biomarker standardisation initiative—IBSI 0.0. 1dev documentation 2019},
  author={Zwanenburg, A and Leger, S and Valli{\`e}res, M and L{\"o}ck, S},
  journal={ProQuest Number: INFORMATION TO ALL USERS},
  year={2021}
}

@article{vanGriethuysen2017radiomics,
  author  = {van Griethuysen, J. J. M. and Fedorov, A. and Parmar, C. and Hosny, A. and Aucoin, N. and Narayan, V. and Beets-Tan, R. G. H. and Fillon-Robin, J. C. and Pieper, S. and Aerts, H. J. W. L.},
  title   = {Computational Radiomics System to Decode the Radiographic Phenotype},
  journal = {Cancer Research},
  year    = {2017},
  volume  = {77},
  number  = {21},
  pages   = {e104--e107},
  doi     = {10.1158/0008-5472.CAN-17-0339}
}

@article{zhang2015ibex,
  title={IBEX: an open infrastructure software platform to facilitate collaborative work in radiomics},
  author={Zhang, Lifei and Fried, David V and Fave, Xenia J and Hunter, Luke A and Yang, Jinzhong and Court, Laurence E},
  journal={Medical physics},
  volume={42},
  number={3},
  pages={1341--1353},
  year={2015},
  publisher={Wiley Online Library}
}

@article{szczypinski2009mazda,
  title={MaZda—a software package for image texture analysis},
  author={Szczypi{\'n}ski, Piotr M and Strzelecki, Micha{\l} and Materka, Andrzej and Klepaczko, Artur},
  journal={Computer methods and programs in biomedicine},
  volume={94},
  number={1},
  pages={66--76},
  year={2009},
  publisher={Elsevier}
}

@article{o2015introduction,
  title={An introduction to convolutional neural networks},
  author={O'shea, Keiron and Nash, Ryan},
  journal={arXiv preprint arXiv:1511.08458},
  year={2015}
}

@article{avanzo2020machine,
  title={Machine and deep learning methods for radiomics},
  author={Avanzo, Michele and Wei, Lise and Stancanello, Joseph and Vallieres, Martin and Rao, Arvind and Morin, Olivier and Mattonen, Sarah A and El Naqa, Issam},
  journal={Medical physics},
  volume={47},
  number={5},
  pages={e185--e202},
  year={2020},
  publisher={Wiley Online Library}
}

@article{varma2006bias,
  title={Bias in error estimation when using cross-validation for model selection},
  author={Varma, Sudhir and Simon, Richard},
  journal={BMC bioinformatics},
  volume={7},
  number={1},
  pages={91},
  year={2006},
  publisher={Springer}
}

@inproceedings{scholkopf1997kernel,
  title={Kernel principal component analysis},
  author={Sch{\"o}lkopf, Bernhard and Smola, Alexander and M{\"u}ller, Klaus-Robert},
  booktitle={International conference on artificial neural networks},
  pages={583--588},
  year={1997},
  organization={Springer}
}

@article{bank2023autoencoders,
  title={Autoencoders},
  author={Bank, Dor and Koenigstein, Noam and Giryes, Raja},
  journal={Machine learning for data science handbook: data mining and knowledge discovery handbook},
  pages={353--374},
  year={2023},
  publisher={Springer}
}

@article{cha1994partial,
  title={Partial least squares},
  author={Cha, Jaesung},
  journal={Adv. Methods Mark. Res},
  volume={407},
  pages={52--78},
  year={1994}
}

@incollection{lee1998independent,
  title={Independent component analysis},
  author={Lee, Te-Won},
  booktitle={Independent component analysis: Theory and applications},
  pages={27--66},
  year={1998},
  publisher={Springer}
}

@article{mackiewicz1993principal,
  title={Principal components analysis (PCA)},
  author={Ma{\'c}kiewicz, Andrzej and Ratajczak, Waldemar},
  journal={Computers \& Geosciences},
  volume={19},
  number={3},
  pages={303--342},
  year={1993},
  publisher={Elsevier}
}

@article{zou2005regularization,
  title={Regularization and variable selection via the elastic net},
  author={Zou, Hui and Hastie, Trevor},
  journal={Journal of the Royal Statistical Society Series B: Statistical Methodology},
  volume={67},
  number={2},
  pages={301--320},
  year={2005},
  publisher={Oxford University Press}
}

@article{tibshirani1996regression,
  title={Regression shrinkage and selection via the lasso},
  author={Tibshirani, Robert},
  journal={Journal of the Royal Statistical Society Series B: Statistical Methodology},
  volume={58},
  number={1},
  pages={267--288},
  year={1996},
  publisher={Oxford University Press}
}

@article{kohavi1997wrappers,
  title={Wrappers for feature subset selection},
  author={Kohavi, Ron and John, George H},
  journal={Artificial intelligence},
  volume={97},
  number={1-2},
  pages={273--324},
  year={1997},
  publisher={Elsevier}
}

@article{peng2005feature,
  title={Feature selection based on mutual information criteria of max-dependency, max-relevance, and min-redundancy},
  author={Peng, Hanchuan and Long, Fuhui and Ding, Chris},
  journal={IEEE Transactions on pattern analysis and machine intelligence},
  volume={27},
  number={8},
  pages={1226--1238},
  year={2005},
  publisher={IEEE}
}

@article{deasy2003cerr,
  title={CERR: a computational environment for radiotherapy research},
  author={Deasy, Joseph O and Blanco, Angel I and Clark, Vanessa H},
  journal={Medical physics},
  volume={30},
  number={5},
  pages={979--985},
  year={2003},
  publisher={Wiley Online Library}
}

@article{ma2025towards,
  title={Towards reliable radiomics modeling: a multi-institutional multi-modality feature repeatability study on head and neck cancer patients},
  author={Ma, Zongrui and others},
  year={2025},
  publisher={Hong Kong Polytechnic University}
}

@article{gillies2016radiomics,
  title={Radiomics: images are more than pictures, they are data},
  author={Gillies, Robert J and Kinahan, Paul E and Hricak, Hedvig},
  journal={Radiology},
  volume={278},
  number={2},
  pages={563--577},
  year={2016},
  publisher={Radiological Society of North America}
}

@article{vial2018role,
  title={The role of deep learning and radiomic feature extraction in cancer-specific predictive modelling: a review},
  author={Vial, Alanna and Stirling, David and Field, Matthew and Ros, Montserrat and Ritz, Christian and Carolan, Martin and Holloway, Lois and Miller, Alexis A},
  journal={Translational Cancer Research},
  volume={7},
  number={3},
  year={2018},
  publisher={AME Publishing Company}
}

@article{mi2020impact,
  title={Impact of different scanners and acquisition parameters on robustness of MR radiomics features based on women’s cervix},
  author={Mi, Honglan and Yuan, Mingyuan and Suo, Shiteng and Cheng, Jiejun and Li, Suqin and Duan, Shaofeng and Lu, Qing},
  journal={Scientific reports},
  volume={10},
  number={1},
  pages={20407},
  year={2020},
  publisher={Nature Publishing Group UK London}
}

@article{joskowicz2019inter,
  title={Inter-observer variability of manual contour delineation of structures in CT},
  author={Joskowicz, Leo and Cohen, D and Caplan, N and Sosna, Jacob},
  journal={European radiology},
  volume={29},
  number={3},
  pages={1391--1399},
  year={2019},
  publisher={Springer}
}

@article{xue2021radiomics,
  title={Radiomics feature reliability assessed by intraclass correlation coefficient: a systematic review},
  author={Xue, Cindy and Yuan, Jing and Lo, Gladys G and Chang, Amy TY and Poon, Darren MC and Wong, Oi Lei and Zhou, Yihang and Chu, Winnie CW},
  journal={Quantitative imaging in medicine and surgery},
  volume={11},
  number={10},
  pages={4431},
  year={2021}
}

@article{hu2023image,
  title={Image harmonization: A review of statistical and deep learning methods for removing batch effects and evaluation metrics for effective harmonization},
  author={Hu, Fengling and Chen, Andrew A and Horng, Hannah and Bashyam, Vishnu and Davatzikos, Christos and Alexander-Bloch, Aaron and Li, Mingyao and Shou, Haochang and Satterthwaite, Theodore D and Yu, Meichen and others},
  journal={NeuroImage},
  volume={274},
  pages={120125},
  year={2023},
  publisher={Elsevier}
}

@article{duron2019gray,
  title={Gray-level discretization impacts reproducible MRI radiomics texture features},
  author={Duron, Lo{\"\i}c and Balvay, Daniel and Vande Perre, Saskia and Bouchouicha, Afef and Savatovsky, Julien and Sadik, Jean-Claude and Thomassin-Naggara, Isabelle and Fournier, Laure and Lecler, Augustin},
  journal={PLoS One},
  volume={14},
  number={3},
  pages={e0213459},
  year={2019},
  publisher={Public Library of Science San Francisco, CA USA}
}

@inproceedings{reinhold2019evaluating,
  title={Evaluating the impact of intensity normalization on {MR} image synthesis},
  author={Reinhold, Jacob C and Dewey, Blake E and Carass, Aaron and Prince, Jerry L},
  booktitle={Medical Imaging 2019: Image Processing},
  volume={10949},
  pages={109493H},
  year={2019},
  organization={International Society for Optics and Photonics}}

@techreport{wang2022critical,
  title={Critical factors in achieving fine-scale functional MRI: Removing sources of inadvertent spatial smoothing},
  author={Wang, Jianbao and Nasr, Shahin and Roe, Anna Wang and Polimeni, Jonathan R},
  year={2022},
  institution={Wiley Online Library}
}

@mastersthesis{sang2024quantifying,
  title={Quantifying Radiomic Texture Characterization Performance on Image Resampling and Discretization},
  author={Sang, Weiwei},
  year={2024},
  school={Duke University}
}

@article{zhao2021understanding,
  title={Understanding sources of variation to improve the reproducibility of radiomics},
  author={Zhao, Binsheng},
  journal={Frontiers in oncology},
  volume={11},
  pages={633176},
  year={2021}
}

@article{zhang2023radiomics,
  title={Radiomics and its feature selection: a review},
  author={Zhang, Wenchao and Guo, Yu and Jin, Qiyu},
  journal={Symmetry},
  volume={15},
  number={10},
  pages={1834},
  year={2023},
  publisher={MDPI}
}

@article{perniciano2025insights,
  title={Insights into radiomics: impact of feature selection and classification},
  author={Perniciano, Alessandra and Loddo, Andrea and Di Ruberto, Cecilia and Pes, Barbara},
  journal={Multimedia Tools and Applications},
  volume={84},
  number={26},
  pages={31695--31721},
  year={2025},
  publisher={Springer}
}

@article{xu2025addressing,
  title={Addressing the current challenges in the clinical application of AI-based Radiomics for cancer imaging},
  author={Xu, Yongzhong and Li, Yunxin and Wang, Feng and Zhang, Yafei and Huang, Delong},
  journal={Frontiers in Medicine},
  volume={12},
  pages={1674397},
  year={2025},
  publisher={Frontiers}
}

@article{linton2025radiomics,
  title={Radiomics in clinical radiology: advances, challenges, and future directions},
  author={Linton-Reid, Kristofer and Chen, Mitchell and Martell, Marc Boubnovski and Posma, Joram M and Aboagye, Eric O},
  journal={Clinical Radiology},
  pages={107165},
  year={2025},
  publisher={Elsevier}
}

@article{castillo2021multi,
  title={A multi-center, multi-vendor study to evaluate the generalizability of a radiomics model for classifying prostate cancer: high grade vs. low grade},
  author={Castillo T, Jose M and Starmans, Martijn PA and Arif, Muhammad and Niessen, Wiro J and Klein, Stefan and Bangma, Chris H and Schoots, Ivo G and Veenland, Jifke F},
  journal={Diagnostics},
  volume={11},
  number={2},
  pages={369},
  year={2021},
  publisher={MDPI}
}

@article{zhou2018radiomics,
  title={Radiomics in brain tumor: image assessment, quantitative feature descriptors, and machine-learning approaches},
  author={Zhou, Mu and Scott, Jacob and Chaudhury, Baishali and Hall, Lawrence and Goldgof, Dmitry and Yeom, Kristen W and Iv, Michael and Ou, Yangming and Kalpathy-Cramer, Jayashree and Napel, Sandy and others},
  journal={American Journal of Neuroradiology},
  volume={39},
  number={2},
  pages={208--216},
  year={2018},
  publisher={American Journal of Neuroradiology}
}

@article{oikonomou2020artificial,
  title={Artificial intelligence in medical imaging: a radiomic guide to precision phenotyping of cardiovascular disease},
  author={Oikonomou, Evangelos K and Siddique, Musib and Antoniades, Charalambos},
  journal={Cardiovascular Research},
  volume={116},
  number={13},
  pages={2040--2054},
  year={2020},
  publisher={Oxford University Press}
}

@article{liu2019applications,
  title={The applications of radiomics in precision diagnosis and treatment of oncology: opportunities and challenges},
  author={Liu, Zhenyu and Wang, Shuo and Dong, Di and Wei, Jingwei and Fang, Cheng and Zhou, Xuezhi and Sun, Kai and Li, Longfei and Li, Bo and Wang, Meiyun and others},
  journal={Theranostics},
  volume={9},
  number={5},
  pages={1303},
  year={2019}
}

@article{sala2017unravelling,
  title={Unravelling tumour heterogeneity using next-generation imaging: radiomics, radiogenomics, and habitat imaging},
  author={Sala, Evis and Mema, E and Himoto, Y and Veeraraghavan, H and Brenton, JD and Snyder, A and Weigelt, B and Vargas, HA},
  journal={Clinical radiology},
  volume={72},
  number={1},
  pages={3--10},
  year={2017},
  publisher={Elsevier}
}

@article{lam2022radiomics,
  title={A radiomics-based machine learning model for prediction of tumor mutational burden in lower-grade gliomas},
  author={Lam, Luu Ho Thanh and Chu, Ngan Thy and Tran, Thi-Oanh and Do, Duyen Thi and Le, Nguyen Quoc Khanh},
  journal={Cancers},
  volume={14},
  number={14},
  pages={3492},
  year={2022},
  publisher={MDPI}
}

@article{kumar2023machine,
  title={Machine-learning-based radiomics for classifying glioma grade from magnetic resonance images of the brain},
  author={Kumar, Anuj and Jha, Ashish Kumar and Agarwal, Jai Prakash and Yadav, Manender and Badhe, Suvarna and Sahay, Ayushi and Epari, Sridhar and Sahu, Arpita and Bhattacharya, Kajari and Chatterjee, Abhishek and others},
  journal={Journal of Personalized Medicine},
  volume={13},
  number={6},
  pages={920},
  year={2023},
  publisher={MDPI}
}

@article{chaddad2025radiomic,
  title={A Radiomic Model for Gliomas Grade and Patient Survival Prediction},
  author={Chaddad, Ahmad and Jia, Pingyue and Hu, Yan and Katib, Yousef and Kateb, Reem and Daqqaq, Tareef Sahal},
  journal={Bioengineering},
  volume={12},
  number={5},
  pages={450},
  year={2025},
  publisher={MDPI}
}

@inproceedings{chaddad2019deep,
  title={Deep radiomic features from MRI scans predict survival outcome of recurrent glioblastoma},
  author={Chaddad, Ahmad and Zhang, Mingli and Desrosiers, Christian and Niazi, Tamim},
  booktitle={International Workshop on Radiomics and Radiogenomics in Neuro-oncology},
  pages={36--43},
  year={2019},
  organization={Springer}
}

@inproceedings{li2019deep,
  title={Deep convolutional neural networks for imaging data based survival analysis of rectal cancer},
  author={Li, Hongming and Boimel, Pamela and Janopaul-Naylor, James and Zhong, Haoyu and Xiao, Ying and Ben-Josef, Edgar and Fan, Yong},
  booktitle={2019 IEEE 16th International Symposium on Biomedical Imaging (ISBI 2019)},
  pages={846--849},
  year={2019},
  organization={IEEE}
}

@article{polidori2023radiomics,
  title={Radiomics applications in cardiac imaging: a comprehensive review},
  author={Polidori, Tiziano and De Santis, Domenico and Rucci, Carlotta and Tremamunno, Giuseppe and Piccinni, Giulia and Pugliese, Luca and Zerunian, Marta and Guido, Gisella and Pucciarelli, Francesco and Bracci, Benedetta and others},
  journal={La radiologia medica},
  volume={128},
  number={8},
  pages={922--933},
  year={2023},
  publisher={Springer}
}

@article{pleshkov2025radiomics,
  title={RADIOMICS IN APPLICATION TO DISEASES OF THE MUSCULOSKELETAL SYSTEM. LITERATURE REVIEW},
  author={Pleshkov, Maksim and Zamyshevskaya, Maria and Kuchinskii, Egor and Jin, Xiance and Zhang, Ji and Zavadovskaya, Vera and Zorkaltsev, Maxim and Kim, Tkhe and Pogonchenkova, Daria and Udodov, Vladimir and others},
  journal={Digital Diagnostics},
  year={2025}
}

@article{lu2018machine,
  title={Machine learning--based radiomics for molecular subtyping of gliomas},
  author={Lu, Chia-Feng and Hsu, Fei-Ting and Hsieh, Kevin Li-Chun and Kao, Yu-Chieh Jill and Cheng, Sho-Jen and Hsu, Justin Bo-Kai and Tsai, Ping-Huei and Chen, Ray-Jade and Huang, Chao-Ching and Yen, Yun and others},
  journal={Clinical Cancer Research},
  volume={24},
  number={18},
  pages={4429--4436},
  year={2018},
  publisher={American Association for Cancer Research}
}

@article{gitto2020mri,
  title={MRI radiomics-based machine-learning classification of bone chondrosarcoma},
  author={Gitto, Salvatore and Cuocolo, Renato and Albano, Domenico and Chianca, Vito and Messina, Carmelo and Gambino, Angelo and Ugga, Lorenzo and Cortese, Maria Cristina and Lazzara, Angelo and Ricci, Domenico and others},
  journal={European Journal of Radiology},
  volume={128},
  pages={109043},
  year={2020},
  publisher={Elsevier}
}

@article{chen2019radiomics,
  title={Radiomics-based machine learning in differentiation between glioblastoma and metastatic brain tumors},
  author={Chen, Chaoyue and Ou, Xuejin and Wang, Jian and Guo, Wen and Ma, Xuelei},
  journal={Frontiers in oncology},
  volume={9},
  pages={806},
  year={2019},
  publisher={Frontiers Media SA}
}

@article{yu2017texture,
  title        = {Texture analysis as a radiomic marker for differentiating renal tumors},
  author       = {Yu, HeiShun and Scalera, Jonathan and Khalid, Maria and Touret, Anne-Sophie and Bloch, Nicolas and Li, Baojun and Qureshi, Muhammad M. and Soto, Jorge A. and Anderson, Stephan W.},
  journal      = {Abdominal Radiology},
  year         = {2017},
  volume       = {42},
  pages        = {2470--2478},
  doi          = {10.1007/s00261-017-1144-1},
  url          = {https://doi.org/10.1007/s00261-017-1144-1}
}

@article{Yi2020CTRadiomicsRCC,
  title        = {Computed Tomography Radiomics for Predicting Pathological Grade of Renal Cell Carcinoma},
  author       = {Yi, Xiaoping and Xiao, Qiao and Zeng, Feiyue and Yin, Hongling and Li, Zan and Qian, Cheng and Wang, Cikui and Lei, Guangwu and Xu, Qingsong and Li, Chuanquan and Li, Minghao and Gong, Guanghui and Zee, Chishing and Guan, Xiao and Liu, Longfei and Chen, Bihong T.},
  journal      = {Frontiers in Oncology},
  volume       = {10},
  pages        = {570396},
  year         = {2020},
  doi          = {10.3389/fonc.2020.570396},
  url          = {https://www.frontiersin.org/journals/oncology/articles/10.3389/fonc.2020.570396/full}
}

@article{Feng2018MachineLearningCT,
  title        = {Machine learning-based quantitative texture analysis of CT images of small renal masses: Differentiation of angiomyolipoma without visible fat from renal cell carcinoma},
  author       = {Feng, Zhichao and Rong, Pengfei and Cao, Peng and Zhou, Qingyu and Zhu, Wenwei and Yan, Zhimin and Liu, Qianyun and Wang, Wei},
  journal      = {European Radiology},
  volume       = {28},
  number       = {4},
  pages        = {1625--1633},
  year         = {2018},
  doi          = {10.1007/s00330-017-5118-z},
  url          = {https://link.springer.com/article/10.1007/s00330-017-5118-z}
}

@article{Deng2020CTTextureRenal,
  title        = {Usefulness of CT texture analysis in differentiating benign and malignant renal tumours},
  author       = {Deng, Y. and others},
  journal      = {Clinical Radiology},
  volume       = {75},
  number       = {2},
  pages        = {108--115},
  year         = {2020},
  doi          = {10.1016/j.crad.2019.09.131},
  url          = {https://www.clinicalradiologyonline.net/article/S0009-9260(19)30573-2/fulltext}
}

@article{Wang2021RadiomicsCT_RCC,
  title        = {Radiomics models based on enhanced computed tomography to distinguish clear cell from non-clear cell renal cell carcinomas},
  author       = {Wang, Ping and Pei, Xu and Yin, Xiao-Ping and Ren, Jia-Liang and Wang, Yun and Ma, Lu-Yao and Du, Xiao-Guang and Gao, Bu-Lang},
  journal      = {Scientific Reports},
  volume       = {11},
  pages        = {13729},
  year         = {2021},
  doi          = {10.1038/s41598-021-93069-z},
  url          = {https://www.nature.com/articles/s41598-021-93069-z}
}

@article{erdim2020prediction,
  title={Prediction of Benign and Malignant Solid Renal Masses: Machine Learning–Based CT Texture Analysis},
  author={Erdim, Cagri and Yardimci, Aytul Hande and Bektas, Ceyda Turan and Kocak, Burak and Baykal Koca, Sevim and Demir, Hale and Kilickesmez, {\"O}zg{\"u}r},
  journal={Academic Radiology},
  volume={27},
  number={10},
  pages={1422--1429},
  year={2020},
  publisher={Elsevier},
  doi={10.1016/j.acra.2019.12.015},
}

@article{alhussaini2022comparative,
  title={Comparative analysis for the distinction of chromophobe renal cell carcinoma from renal oncocytoma in computed tomography imaging using machine learning radiomics analysis},
  author={Alhussaini, Abeer J and Steele, J Douglas and Nabi, Ghulam},
  journal={Cancers},
  volume={14},
  number={15},
  pages={3609},
  year={2022},
  publisher={MDPI}
}

@article{kilicarslan2025fusion,
  title={Fusion-Based Deep Learning Approach for Renal Cell Carcinoma Subtype Detection Using Multi-Phasic MRI Data},
  author={Kilicarslan, Gulhan and Cetintas, Dilber and Tuncer, Taner and Yildirim, Muhammed},
  journal={Diagnostics},
  volume={15},
  number={13},
  pages={1636},
  year={2025},
  publisher={MDPI}
}

@article{He2023DeepRadiomicsCysticRenal,
  title        = {Deep learning and radiomic feature-based blending ensemble classifier for malignancy risk prediction in cystic renal lesions},
  author       = {He, Quan-Hao and Feng, Jia-Jun and Lv, Fa-Jin and Jiang, Qing and Xiao, Ming-Zhao},
  journal      = {Insights into Imaging},
  volume       = {14},
  number       = {1},
  pages        = {6},
  year         = {2023},
  doi          = {10.1186/s13244-022-01349-7},
  url          = {https://link.springer.com/article/10.1186/s13244-022-01349-7}
}

@article{Magnuska2024Radiol232554,
  title        = {Combining Radiomics and Autoencoders to Distinguish Benign and Malignant Tumors},
  author       = {Magnuska, Zuzanna Anna and Roy, Rijo and Palmowski, Moritz and Kohlen, Matthias and Winkler, Brigitte Sophia and Pfeil, Tatjana and Boor, Peter and Schulz, Volkmar and Krauss, Katja and Stickeler, Elmar},
  journal      = {Radiology},
  volume       = {312},
  number       = {3},
  pages        = {e232554},
  year         = {2024},
  doi          = {10.1148/radiol.232554},
  url          = {https://pubs.rsna.org/doi/10.1148/radiol.232554}
}

@inproceedings{xu2023classification,
  title={Classification of Benign and malignant renal tumors based on CT scans and Clinical Data using machine learning methods},
  author={Xu, Jie and He, Xing and Shao, Wei and Bian, Jiang and Terry, Russell},
  booktitle={Informatics},
  volume={10},
  number={3},
  pages={55},
  year={2023},
  organization={MDPI}
}

@article{nie2020ct,
  title={A CT-based radiomics nomogram for differentiation of renal angiomyolipoma without visible fat from homogeneous clear cell renal cell carcinoma},
  author={Nie, Pei and Yang, Guangjie and Wang, Zhenguang and Yan, Lei and Miao, Wenjie and Hao, Dapeng and Wu, Jie and Zhao, Yujun and Gong, Aidi and Cui, Jingjing and others},
  journal={European radiology},
  volume={30},
  number={2},
  pages={1274--1284},
  year={2020},
  publisher={Springer}
}

@article{sun2020radiologic,
  title={Radiologic-radiomic machine learning models for differentiation of benign and malignant solid renal masses: comparison with expert-level radiologists},
  author={Sun, Xue-Ying and Feng, Qiu-Xia and Xu, Xun and Zhang, Jing and Zhu, Fei-Peng and Yang, Yan-Hao and Zhang, Yu-Dong},
  journal={American Journal of Roentgenology},
  volume={214},
  number={1},
  pages={W44--W54},
  year={2020},
  publisher={American Roentgen Ray Society}
}

@article{uhlig2020discriminating,
  title={Discriminating malignant and benign clinical T1 renal masses on computed tomography: A pragmatic radiomics and machine learning approach},
  author={Uhlig, Johannes and Biggemann, Lorenz and Nietert, Manuel M and Bei{\ss}barth, Tim and Lotz, Joachim and Kim, Hyun S and Trojan, Lutz and Uhlig, Annemarie},
  journal={Medicine},
  volume={99},
  number={16},
  pages={e19725},
  year={2020},
  publisher={LWW}
}

@article{gurbani2021evaluation,
  title={Evaluation of radiomics and machine learning in identification of aggressive tumor features in renal cell carcinoma (RCC)},
  author={Gurbani, Sidharth and Morgan, Dane and Jog, Varun and Dreyfuss, Leo and Shen, Mingren and Das, Arighno and Abel, E Jason and Lubner, Meghan G},
  journal={Abdominal Radiology},
  volume={46},
  number={9},
  pages={4278--4288},
  year={2021},
  publisher={Springer}
}

\end{document}